\documentclass[aps,twocolumn,showpacs,superscriptaddress, amsmath,amssymb]{revtex4}

\usepackage{graphicx}
\usepackage{dcolumn}
\usepackage{bm}
\usepackage{graphics}
\usepackage{subfigure}
\usepackage{amsmath}
\usepackage[T1]{fontenc}
\usepackage{color}
\usepackage{bm}

\begin{document}

\title{Micromechanical description of the compaction of soft pentagon assemblies}
\author{Manuel C\'ardenas-Barrantes}
\email{manuel-antonio.cardenas-barrantes@umontpellier.fr}
\affiliation{LMGC, Universit\'e de Montpellier, CNRS, Montpellier, France}
\author{David Cantor}
\email{david.cantor@polymtl.ca}
\affiliation{Department of Civil, Geological and Mining Engineering, Polytechnique Montr\'eal, Qu\'ebec, Canada}
\author{Jonathan Bar\'es}
\email{jonathan.bares@umontpellier.fr}
\affiliation{LMGC, Universit\'e de Montpellier, CNRS, Montpellier, France}
\author{Mathieu Renouf}
\email{mathieu.renouf@umontpellier.fr}
\affiliation{LMGC, Universit\'e de Montpellier, CNRS, Montpellier, France}
\author{Emilien Az\'ema}
\email{emilien.azema@umontpellier.fr}
\affiliation{LMGC, Universit\'e de Montpellier, CNRS, Montpellier, France}
\affiliation{Institut Universitaire de France (IUF), Paris, France}

\begin{abstract}
We analyze the isotropic compaction of assemblies composed
of soft pentagons interacting through classical Coulomb friction via numerical simulations.
The effect of the initial particle shape is discussed by comparing packings of pentagons
with packings of soft circular particles. We characterize the evolution of the packing fraction, the elastic modulus, and the microstructure (particle rearrangement, connectivity, contact force and particle stress distributions) as a function of the applied stresses. Both systems behave similarly;  the packing fraction increases and tends asymptotically to a maximum value $\phi_{max}$, where the bulk modulus diverges. At the microscopic scale we show that particle rearrangements occur even beyond the jammed state,
the mean coordination increases as a square root of the packing fraction and, the force and stress distributions become
more homogeneous as the packing fraction increases.
Soft pentagons present larger particle rearrangements than circular ones, and such behavior decreases proportionally to the friction. Interestingly, the friction between particles also contributes to a better
homogenization of the contact force network in both systems.
From the expression of the granular stress tensor, we develop a model that describes the compaction behavior as a function
of the applied pressure, the Young modulus and the initial shape of the particles.
This model, settled on the joint evolution of the particle connectivity and the contact stress, provides outstanding predictions from the jamming point up to very high densities.
\end{abstract}

\maketitle

\section{Introduction}
Matter composed of soft particles covers a wide class of materials such as emulsions, microgels,
foams \cite{Katgert2010_Jamming,Katgert2019_The}, biological cells \cite{Maurin2008_Mechanical,Chelin2013_Simulation,Bi2016_prx,Mauer2018} and
metal powders \cite{Heckel1961,Parilak2017,Montes2018}.
Under compression, the common feature of such materials is related to their ability to deform rather than to break. They can also fill in a better way the void space than the rigid particle assemblies do. Such remarkable properties, essentially due to
the shape change, make an important distinction with rigid particle systems. In particular, since the particle deformation is generally governed by a characteristic stress (e.g. the Young modulus), the behavior of soft particles assemblies are much more sensitive to the confining pressure than rigid particles assemblies.

The numerical and experimental investigation of particulate packings composed of soft, highly deformable particles is an extensive and promising field of study. This is not only due to the development of advanced experimental devices or the increasing computational power and numerical technics, but because many fundamental behaviors have already been described for assemblies of rigid grains \cite{Andreotti2013,guyon2020built}.
Among these behaviors, we find the transition to jamming \cite{Liu1998_Jamming,Silvert2002_Analogies,Corwin2005_Structural,Keys2007,Majmudar2007_Jamming},
the force transmission \cite{Bathurst1988,Rothenburg1989_Analytical,Radjai1998_Bimodal,Kruyt2014_On},
or the effects induced by particle geometry \cite{Nguyen2014_Effect,Wiacek2014_Effect,Athanassiadis2014_Particle,Cantor2020_Microstructural,Oquendo2020_Densest,Donev2007_Underconstrained,Azema2013_Packings,Zhao2016_Packings,Nie2010_Effect,Landauer2020_Particle}, to name a few.

An open question today is how robust these findings are with respect to soft particle assemblies?
On the one hand, one of the main difficulties with experiments is to make relevant quantitative measurements at the grain
scale, for example, to follow the shape change of the particles \cite{Katgert2010_Jamming,Vu2019a}.
On the other hand, introducing the correct particle shape deformations in numerical simulations with discrete element methods rises various technical difficulties, mainly seen in the increase of the computational time in the simulations. In this direction different discrete element strategies have been proposed, such as the Bonded-Particle Method \cite{Nezamaba2016_Modeling,Rojek2018_The} and couplings between classical finite element or mesh-less methods \cite{Munjiza_2004,Boromand2018,Jonsson2019_Evaluation,Nezamabadi2019,Wang2020_Particulate,Latham2020_A}.
The latter methods, although computational expensive, have the advantage of closely represent the geometry of the particles.

At the macroscale, the compaction beyond the jamming point is a subject of great interest.
A large number of compaction equations have been proposed \cite{Heckel1961,Carroll1984,Panelli2001,Secondi2002_Modelling,Montes2018,Zhang2014,Parilak2017}, but only recently a micro-mechanical-based model for circular particle assemblies has been established \cite{Cantor2020_Compaction}.
At the microscopic scale, by studying the behavior of foams and others assembly of rubber-like particles,
it has been shown that the mean coordination number increases as a square root of the packing fraction \cite{Durian1995_Foam,Majmudar2007_Jamming,Katgert2010_Jamming,Vu2020_compaction,Vu2021_Effects}, also
the stress distribution seems to become more homogeneous as the packing fraction increases \cite{Katgert2019_The}.
But, in general terms, the literature on this subject is still in its beginnings.

In this paper, we analyze the compaction behavior of assemblies composed of soft pentagons, beyond the jamming point, by means of a coupled discrete element and finite element method: the non-smooth contact dynamics (NSCD) approach.
In order to evidence the effects of the initial shape on the macro and microstructure, assemblies of circular particles under similar characteristics of compression are simulated. The contact friction is also systematically studied.

The paper is structured as follows. The numerical procedures and the compression test are described in Sec. \ref{num}.
Section \ref{Sec_macro} is focused on the evolution of the packing fraction and the bulk properties
beyond the jamming as a function of the applied pressure and for different values of friction.
The microstructure, described in terms of particle mobility, connectivity, force and stress transmission,
is discussed in Sec. \ref{SecMicro}. In Sec. \ref{micro-meca-sec}, we present the
micro-structural elements behind the evolution of the packing fraction and the bulk properties at the macroscopic scale.
Finally, conclusions and perspectives are discussed in Section \ref{conclu}.

\section{Numerical Procedure}
\label{num}
In the NSCD's frame \cite{Jean1999,Dubois2018}, we simulate packings of
deformable and incompressible particles with two different shapes,
regular pentagons and disks, under external
isotropic compression. The NSCD method consists in coupling the finite elements method (FEM), used to model the
deformable particle itself, and the contact dynamics (CD) method, used to deal with the dynamics of the particles and the contact forces
within the packing. The CD method considers a contact law with non-penetrability, and no regularization of the friction
law between the particles for the determination of contact forces.
The deformable particles are modeled following a neo-Hookean incompressible material law.
The simulations are implemented on the open-source platform LMGC90 \cite{LMGC90_web}, where the NCSD algorithm is parallelized \cite{Renouf2004}.

\begin{figure}
\centering
\includegraphics[width=0.44\linewidth]{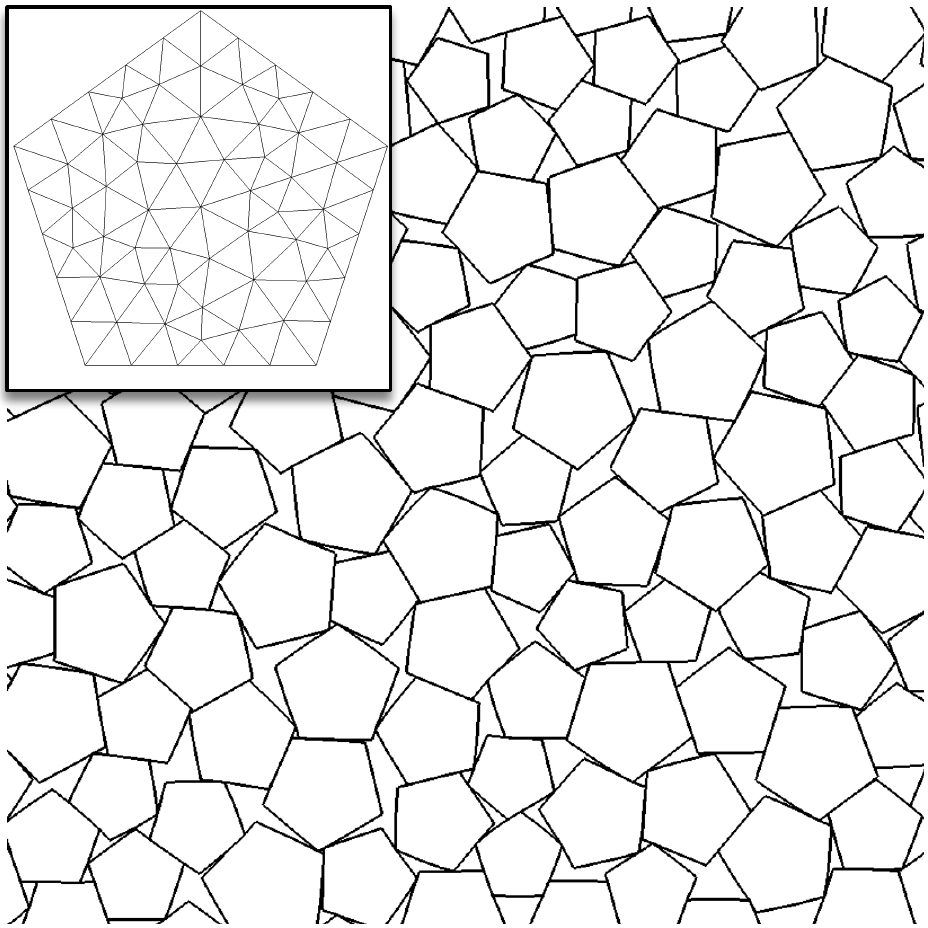}(a)
\includegraphics[width=0.44\linewidth]{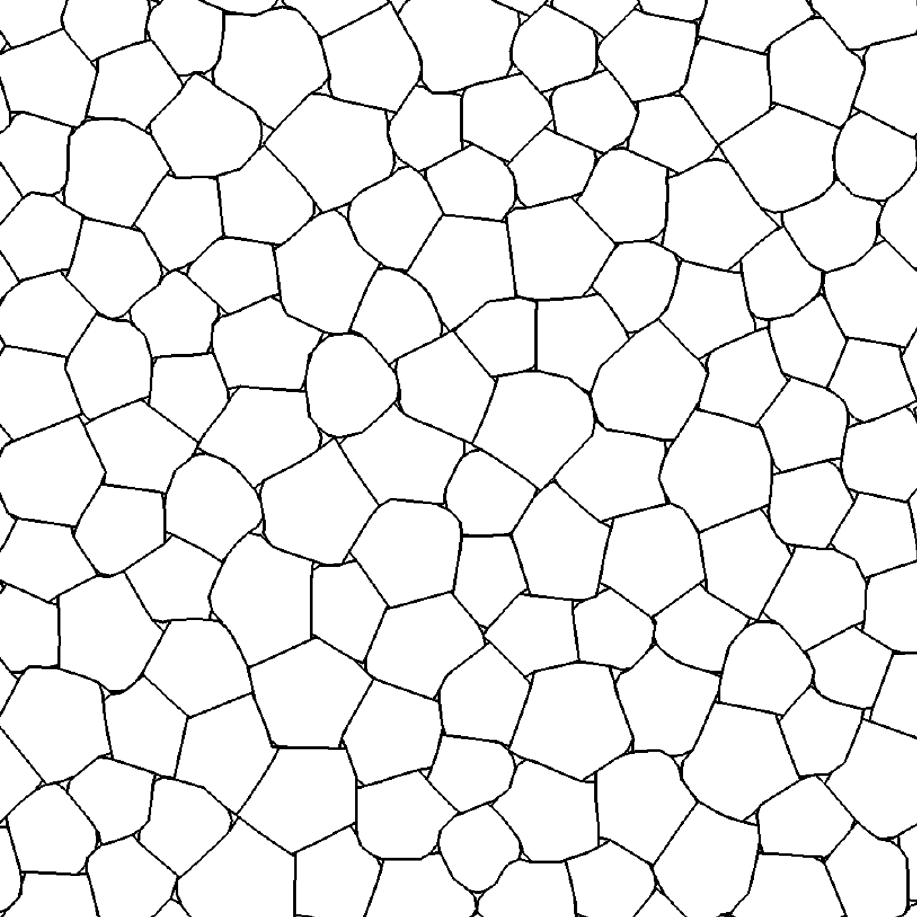}(b)
\includegraphics[width=0.44\linewidth]{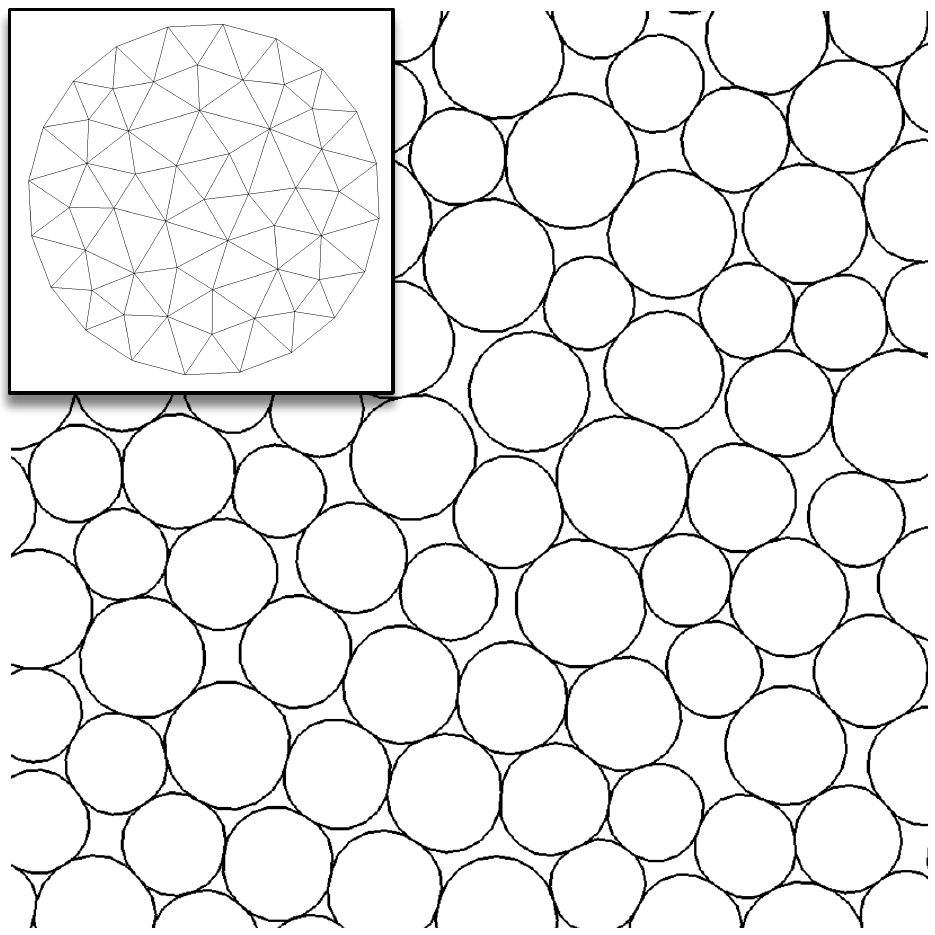}(c)
\includegraphics[width=0.44\linewidth]{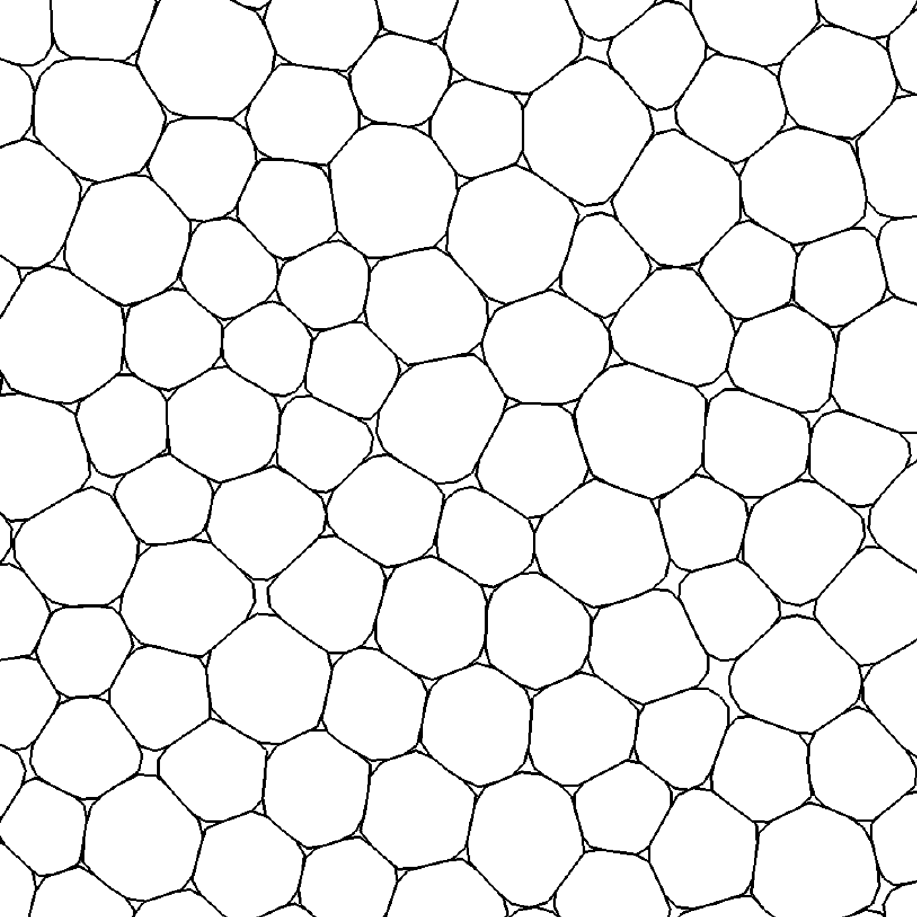}(d)
\caption{Close-up views of the assembly of pentagons (a,b) and disks (c,d) at $\mu=0$ and
for $P/E = 0.025$ (a,c) and $P/E = 0.3$ (b,d).
The insets of (a) and (c) show the finite element mesh used for both pentagons and disks, respectively. }
 \label{fig:Snap-shots}
 \end{figure}

First, $N_p = 500$ rigid particles are randomly placed into a
two-dimentional square box bounded by rigid walls and isotropically compressed under a very small stress $\sigma_0$ up to its jamming point.
At this point, the box size is $L_0\times L_0$.
To avoid crystallization, a small grain size polydispersity around the mean diameter of the
particles $\left<d\right>$ is introduced ($ d \in \left[0.8 \left<d \right>,1.2 \left<d \right> \right]$).
For the case of a pentagon, $d$ is the diameter of its circumcircle.
Second, the particles are meshed with $92$ triangular elements (inset in Fig. \ref{fig:Snap-shots}).
We use a constant Poisson's ratio equals to $0.495$ and a Young modulus $E$ such that $\sigma_0/E<<1$.

Finally, the packings are isotropically compressed imposing the same constant velocity $v$ on the
boundaries of the box.
The velocity $v$ is carefully chosen to be sure that the system is always in the quasi-static regime defined by the inertial number $I = \dot \gamma \langle d \rangle \sqrt{{\rho/P}}$, with $\dot \gamma = v/L_0$, $\rho$ the particle density
and $P$ the confining stress \cite{GDRMiDi2004_On,Andreotti2013}.
The quasi-static limit is ensured for $I<<1$. Thus, in all the simulations, $v$ is computed from the inertia parameter
replacing $I$ by $I_0=10^{-4}$ and $P$ by $\sigma_0$. In this way, the inertia parameters remains below $10^{-4}$ during the compaction process.
We performed a large number of tests for different coefficient of friction between particles,
$\mu \in \left[0.0, 0.1, 0.2, 0.4, 0.8 \right]$, while we kept the coefficient of friction with the walls equal to zero. The
gravity is set to zero.
Figure \ref{fig:Snap-shots} shows frictionless assemblies of pentagons and disks at the jammed state and beyond.

\section{Macroscopic behavior}
\label{Sec_macro}

\subsection{Macroscopic variables}
Under isotropic compression, the confining stress acting on the assembly is given by $P=F/L$, where $F$ is
the computed force on the walls and $L$ its length. For a granular assembly,
we can also compute the confining stress from the granular stress tensor $\bm \sigma$. To do so, we start by computing the tensorial moment ${\bm M}^i$ of each particle $i$,  defined as \cite{Andreotti2013}:
\begin{equation}
M^i_{\alpha \beta} = \sum_{c \in i} f_{\alpha}^c r_{\beta}^c,
\label{eq:M}
\end{equation}
where  $f_{\alpha}^c$ is the $\alpha$ component of the force acting on the
particle $i$ at the contact $c$, $r_{\beta}^c$ is the $\beta$-th component
of the position vector of the same contact $c$, and the sum
runs over the contacts of the particle i ($c \in i$).
Then, the average stress tensor $\bm \sigma $, in the volume $V$, is given by:
\begin{equation}
{\bm \sigma } = \frac{1}{V} \sum_{i \in V} {\bm M}^i.
\label{eq:sigma}
\end{equation}
The confining stress then takes the form $P = (\sigma_1+\sigma_2)/2$, where $\sigma_1=\sigma_2$ are the principal stress values of $\bm \sigma$.

From a geometrical point of view, the cumulative volumetric strain is defined by
\begin{equation}
\varepsilon = -\ln \frac{\phi_0}{\phi},
\label{eq:volum_phi}
\end{equation}
with the packing fraction $\phi$ defined as $\phi = \sum_{i\in V} V_i /V$, where $V_i$ is the volume of the particle $i$. Here, $\phi_0$ is the initial packing fraction of the assembly (i.e., at $V=V_0$).

\subsection{Evolution of the packing fraction and bulk modulus}
\begin{figure}
\centering
\includegraphics[width=\linewidth]{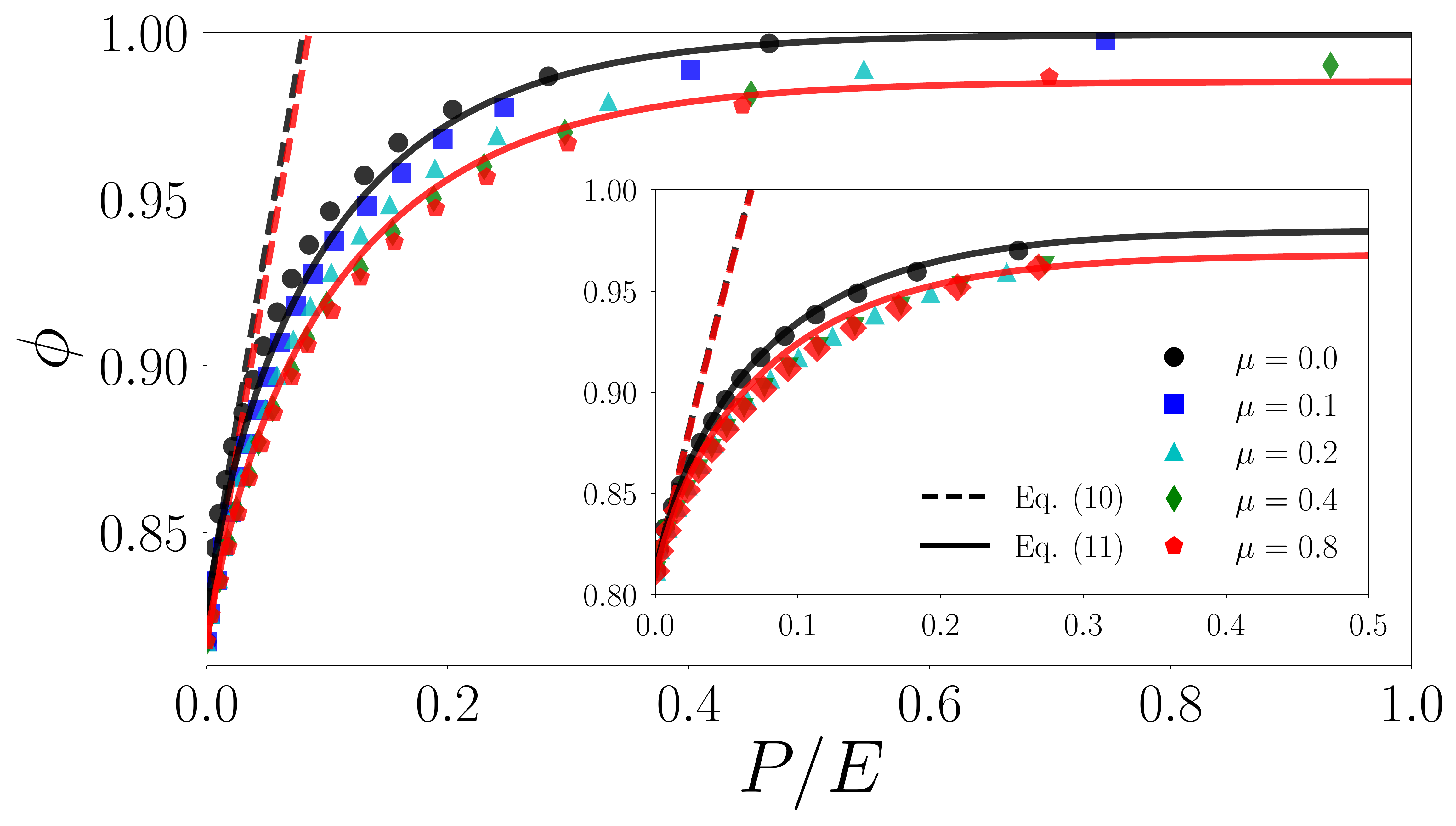}(a)
\includegraphics[width=\linewidth]{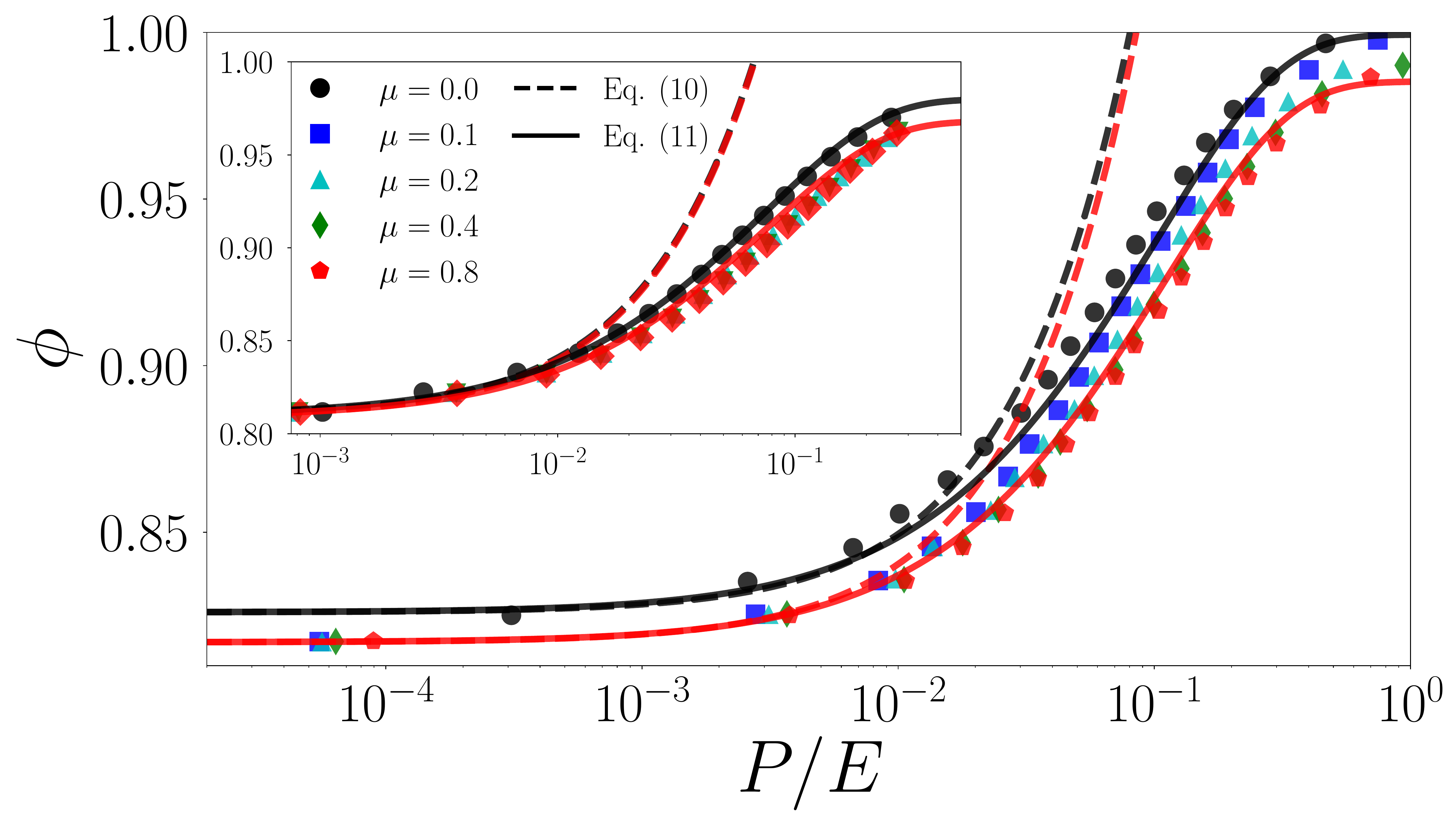}(b)
\caption{(Color online)
Packing fraction $\phi$ as a function of the reduced pression $P/E$ for assemblies of soft pentagons, and different values of friction, in (a) lin-lin scale and (b) log-lin scale. The insets show the same data for assemblies composed of soft disks.
The dashed lines are the elastic approximations (Eq. (\ref{eq:Pgsd})) and the continuous lines are the predictions given by Eq. (\ref{eq:Pglobal}) for $\mu=0$ (black) and $\mu=0.8$ (red).}
\label{fig:pressure-phi}
\end{figure}

Figure \ref{fig:pressure-phi} shows the evolution of $\phi$ as a function of the reduced pressure
$P/E$ for assemblies of pentagons and disks (insets).
Regardless of the value of friction, the packing fraction follows the same general trends.
From the jammed state, the packing fraction increases rapidly for small values of $P/E$, and then, reaches an asymptote at $\phi_{max}$.
We observe a progressive separation of the compaction curves as the friction increases, which results in a slight
decrease in the value of $\phi_{max}$. This offset between the curves is more evident in the case of pentagons than in the case of disks.
Such difference is explained by the fact that the friction and the particle shape restrict the mobility of the particles.
This point will be discussed in detail in Sec. \ref{SecMicro}.

From the $\phi-P/E$ relation, it is also possible to characterize the bulk properties of the assemblies through the definition of the bulk modulus $K$:
\begin{equation}
\label{Eq_Modulus}
K(\phi)= \frac{dP}{d\phi} \cdot \frac{d\phi}{d\varepsilon}.
\end{equation}

\begin{figure}
\centering
\includegraphics[width=\linewidth]{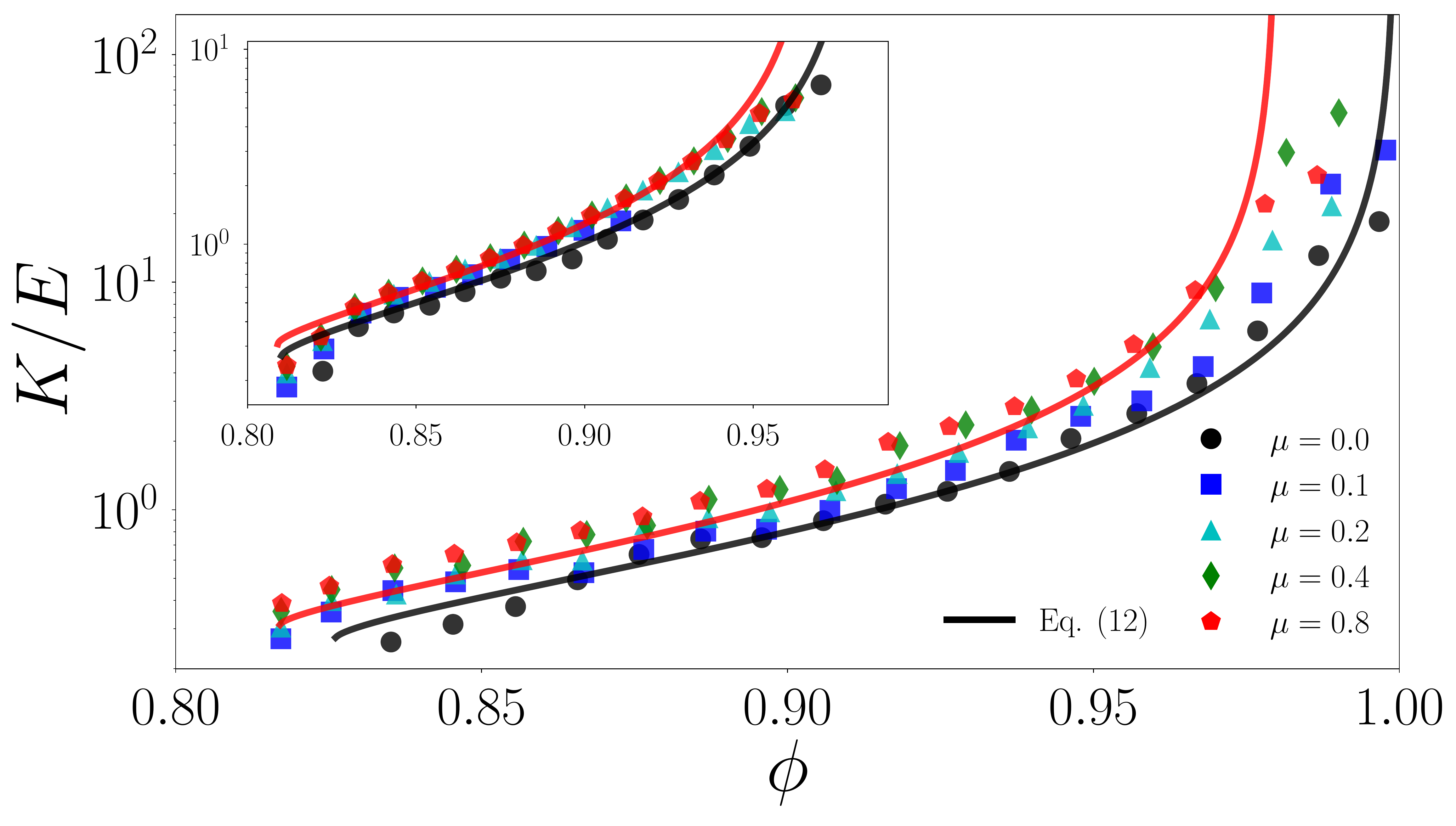}
\caption{(Color online) Evolution of the bulk modulus $K$ normalized by E as a
function of the packing fraction $\phi$ for soft assemblies of pentagons and disks (inset).
The predictions given by Eq. (\ref{eq:Kglobal}) are shown in continuous lines.}
\label{fig:K-phi}
\end{figure}

Figure \ref{fig:K-phi} shows the evolution of $K(\phi)$ as a function of $\phi$ in assemblies
of pentagons and disks (inset). We observe that the bulk evolution follows the same trends, regardless of the shape
of the particles and friction. $K$ appears to be an increasing function of $\phi$, with a divergence at $\phi_{max}$. This divergence is expected, since the system tends to oppose its own compression due to the progressive filling of the void space and the intrinsic incompressible behavior of the particles.
In other words, the assembly of soft particles begins to behave as a rigid body.

In this section we observe that, at the macroscale, the compaction behavior beyond the jamming of assembly of soft
pentagons and soft disks is similar.
Small differences appear mainly on the values of the maximum packing fraction that each system
can reach, which mainly depends on the friction.

\section{Microstructural aspects}
\label{SecMicro}

\subsection{Geometrical features}
\subsubsection{Particle rearrangements}

A way to quantify the rearrangement of an assembly of particles is to measure the deviation of the particles movement from a
reference direction where the reference direction can be defined from the ideal case of a continuous, homogeneous and deformable medium. The difference between the actual displacement and this referential displacement is the so called non-affine motion.
During compression, each material point would move, on average, towards the geometric center of the system.
Thus, we define a rearrangement parameter for each particle $i$, denoted by $\hat{\theta}_i$, as the absolute value of the angle, $\theta_i$, between its velocity $\bm v_i$ and the vector defined from its center and the geometric center of the assembly, divided by $\pi$ ($\hat{\theta}_i=|\theta_i|/\pi$).

\begin{figure}
\centering
\includegraphics[width=\linewidth]{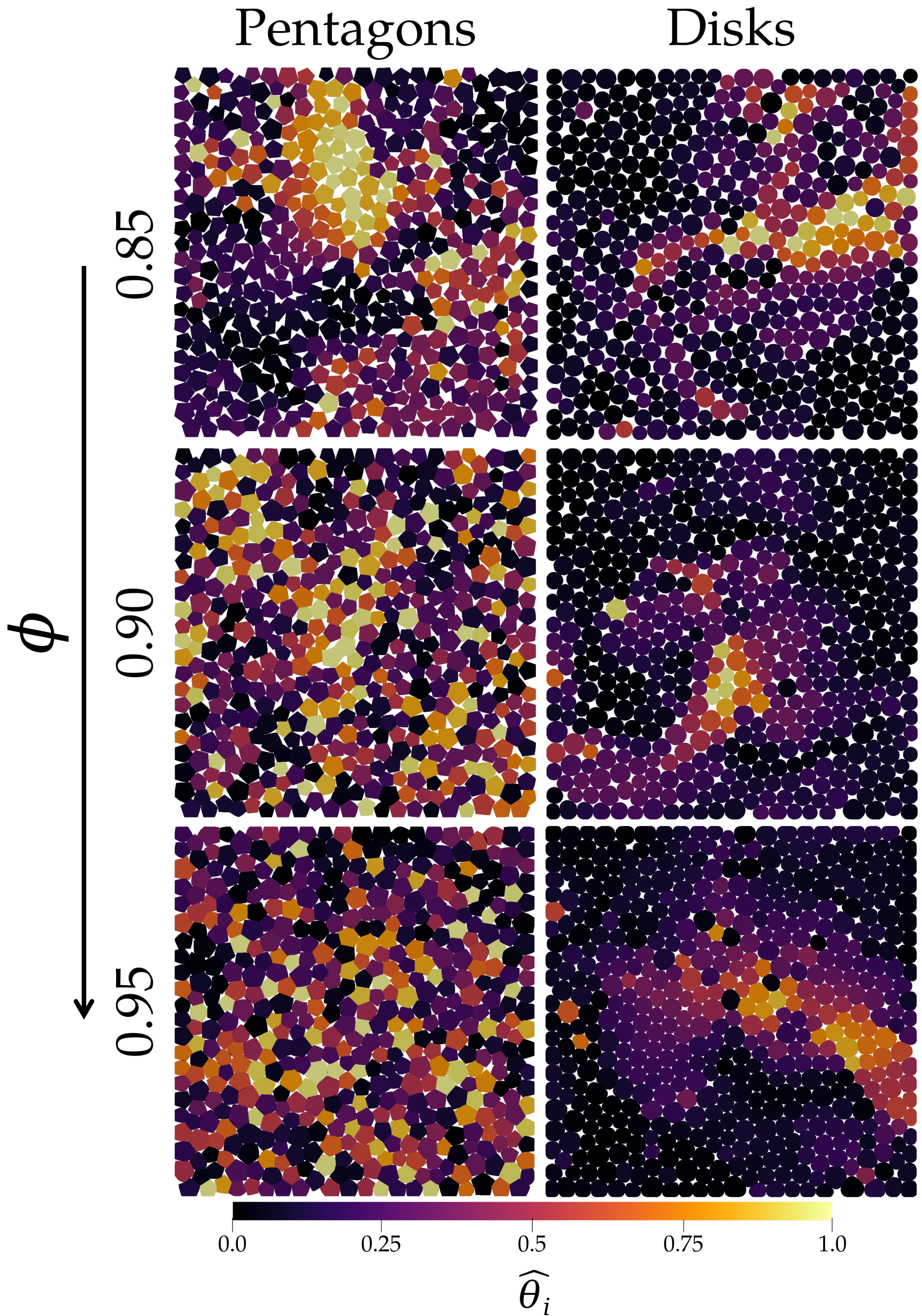}
\caption{(Color online)  Color map of the particles rearrangement parameter $\hat{\theta}_i$ in assemblies of frictionless disks and pentagons for different levels of compaction.}
\label{fig:rearreng}
\end{figure}

Figure \ref{fig:rearreng} shows a color map of the particle rearrangement parameter for each particle at three different levels of compaction.
First, it can be seen that the rearrangement of the particles is highly inhomogeneous during the compaction.
Secondly, we observe zones where particles are rearranged as clusters \cite{bares2017_Local}.

\begin{figure}
\centering
\includegraphics[width=\linewidth]{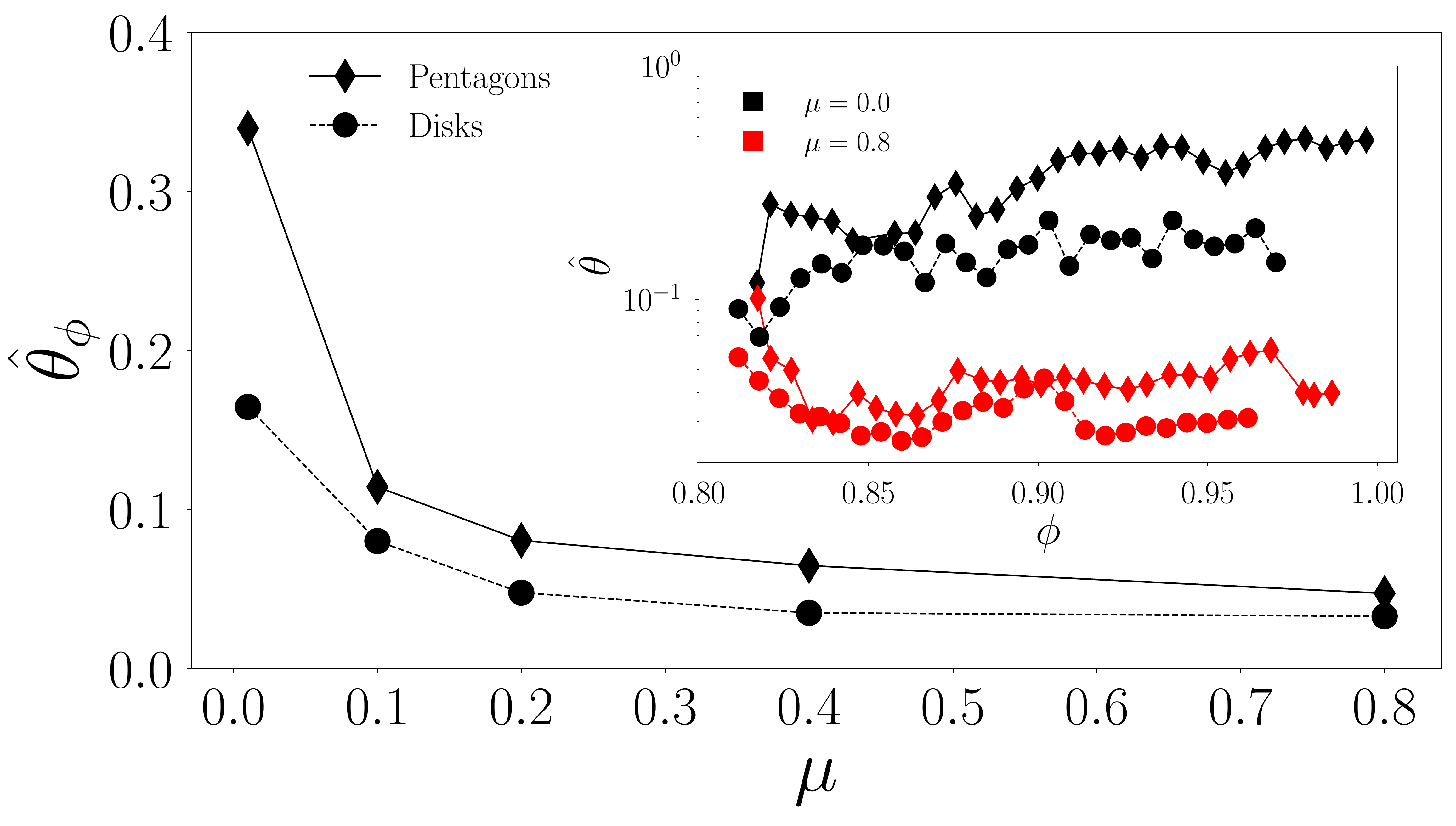}
\caption{(Color online) $\hat{\theta}_{\phi}$ as a function of the friction coefficient for assemblies of pentagons (continuous line) and disks (dashed line).
The inset shows the evolution of $\hat{\theta}$ as a function of the packing fraction $\phi$ for two different values of friction.}
\label{fig:rearreng_mean_theta}
\end{figure}

In the inset of Fig. \ref{fig:rearreng_mean_theta} we see the evolution of the mean value of the rearrangement parameter $\hat{\theta} = \langle\hat\theta_i\rangle_i$ as a function of the packing fraction, for $\mu=0$ and $\mu=0.8$.
Basically, for low friction coefficient, $\hat{\theta}$ slowly increases with $\phi$, while it decreases for larger friction coefficients.
In other words, the particle rearrangements, although small, occurs even after the jamming state and at each stage of the deformation.

To have a better idea of the reorganization of the particles along the compaction process, we compute $\hat{\theta}_{\phi}$,
the asymptotic value of $\hat{\theta}$ as the packing fraction goes to $\phi_{max}$, and plot it as a function of the friction coefficient in Fig. \ref{fig:rearreng_mean_theta}.
We see that low friction allows larger particle rearrangements while high friction tends to prevent it.
Another point to note is that the particle reorganization is higher in pentagon assemblies. In this case, sliding is enhanced by side-side contact as previously shown in non-deformable particle assemblies \cite{Azema2007,Azema2013_Packings}.
From these measures, we can deduce that the differences in the evolution of the compaction curves observed in Fig. \ref{fig:pressure-phi}
are related to the small local rearrangements in the system. This observation also highlights the irreversible
nature of the compaction beyond the jamming, confirming recent works \cite{Vu2020_compaction,Vu2021_Effects}.

\subsubsection{Shape parameter}
\begin{figure}
\centering
\includegraphics[width=\linewidth]{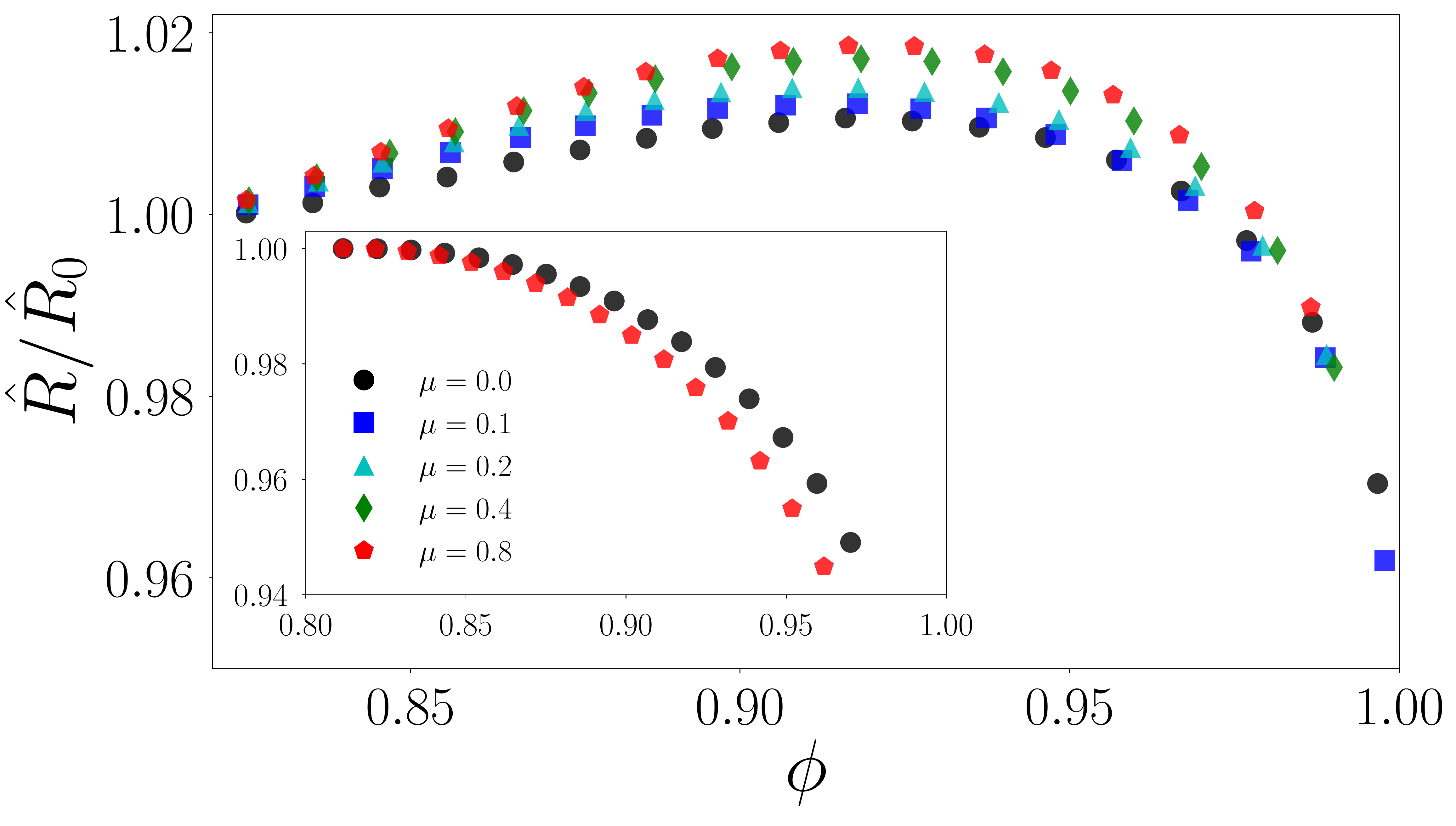}(a)
\includegraphics[width=\linewidth]{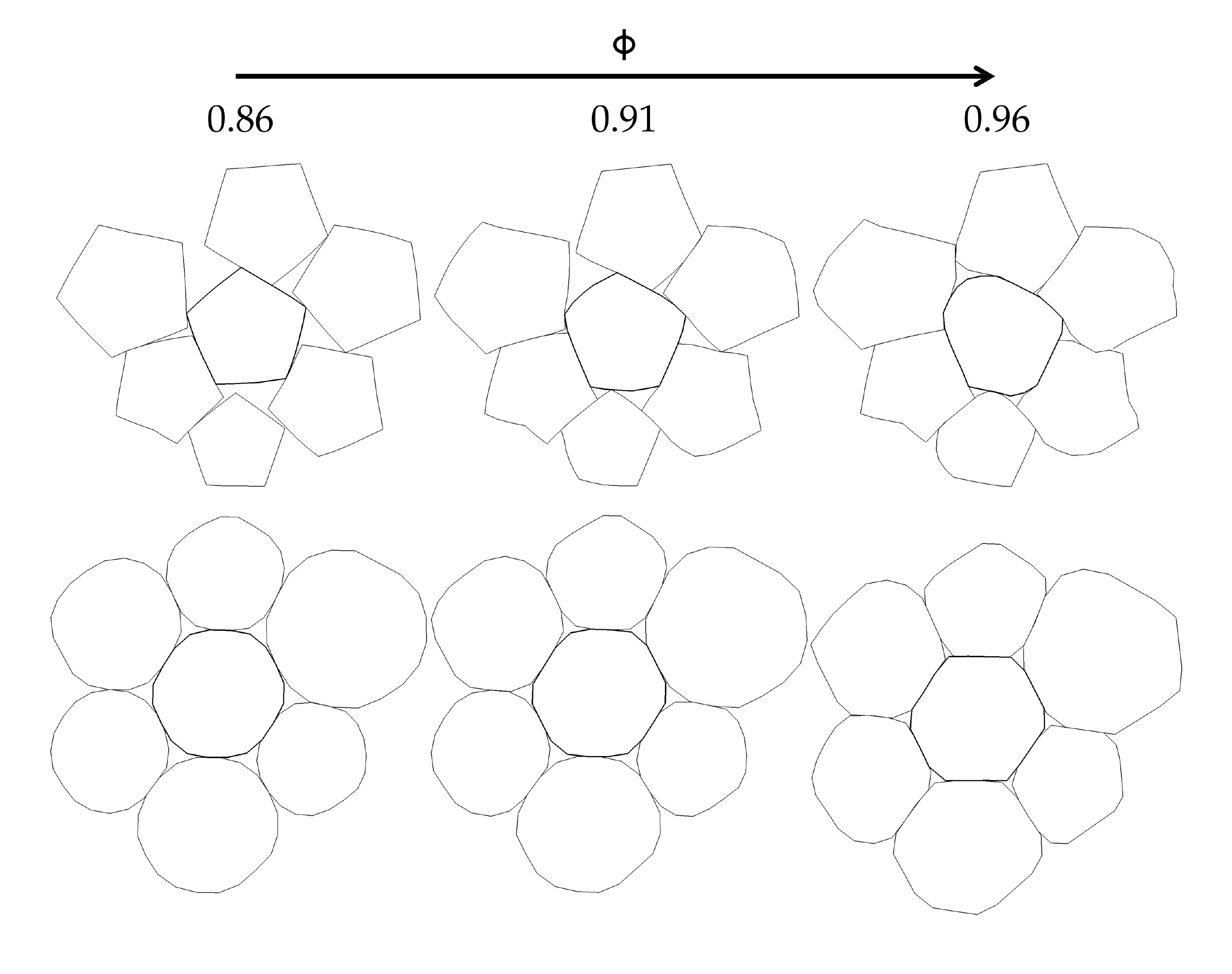}(b)
\caption{(Color online) (a) Evolution of $\hat{R}/\hat{R}_0$ as a function of $\phi$ for different values of friction in assemblies of pentagons and disks (inset).
(b) Group of particles extracted from the assemblies of pentagons and disks, respectively, undergoing the same cumulative packing fraction.}
\label{fig:shape_changing}
\end{figure}

Besides the particle rearrangement, the mechanical properties of soft particle assemblies are intrinsically related to the local deformations of the particles. Let us define the circularity index as usual,
\begin{equation}
\hat{R} = \left\langle 4\pi \frac{V_i}{a_i^2} \right\rangle_i,
\end{equation}
with $a_i$ the particle perimeter and $\langle ...\rangle_i$ the average over the particles in the volume $V$.
Fig. \ref{fig:shape_changing}(a) shows the evolution of $\hat{R}$ scaled with the initial circularity at the jammed state, $\hat{R}_ 0$,  as a function of $\phi$ for different values of the friction.
For disks, $\hat{R}_0$ is equal to $1$, and $\hat{R}/\hat{R}_0$ decreases as the packing fraction increases.
As shown in Fig. \ref{fig:shape_changing}(b),
the disks turn progressively into non-regular polygonal shapes with rounded corners.
In contrast, for pentagons, $\hat{R}/\hat{R}_0$ increases with $\phi$, with $\hat{R}_0 \simeq 0.86$ (the value of the circularity for regular pentagons), to a maximum friction-dependent value and quickly decreases, regardless of the friction.
Initially, as illustrated in Fig. \ref{fig:shape_changing}(b), the pentagons tend to adopt a rounded and regular shape by smoothing the corners. However, beyond a maximal circularity value, a non-regular polygonal shape is observed. This, along with intruding-corner effects into free space, are some of the facts that contributes to pentagons to achieve higher packing fraction compared to assemblies of disks.

\subsection{Particle connectivity}
The first statistical quantity to describe the contact network is the coordination number $Z$, defined as the average number of contacts per particle for non-rattler ones.
At the jammed state, the packing structure is characterized by a minimal value $Z_0$, which depends on the friction coefficient and the packing preparation. For instance, in frictionless assemblies of disks, $Z_0=4$, while $Z_0\in[3,4]$ if friction is activated \cite{Moukarzel1998_Isostatic,Roux2000_Geometric,Hecke2009_Jamming}.
For assemblies of rigid polygons, the jammed-state coordination number $Z_0$, remains close to 4, regardless of the value of the friction \cite{Nguyen2014_Effect,Zhao2019_Jamming,xu2017_Jamming}.

\begin{figure}
\centering
\includegraphics[width=\linewidth]{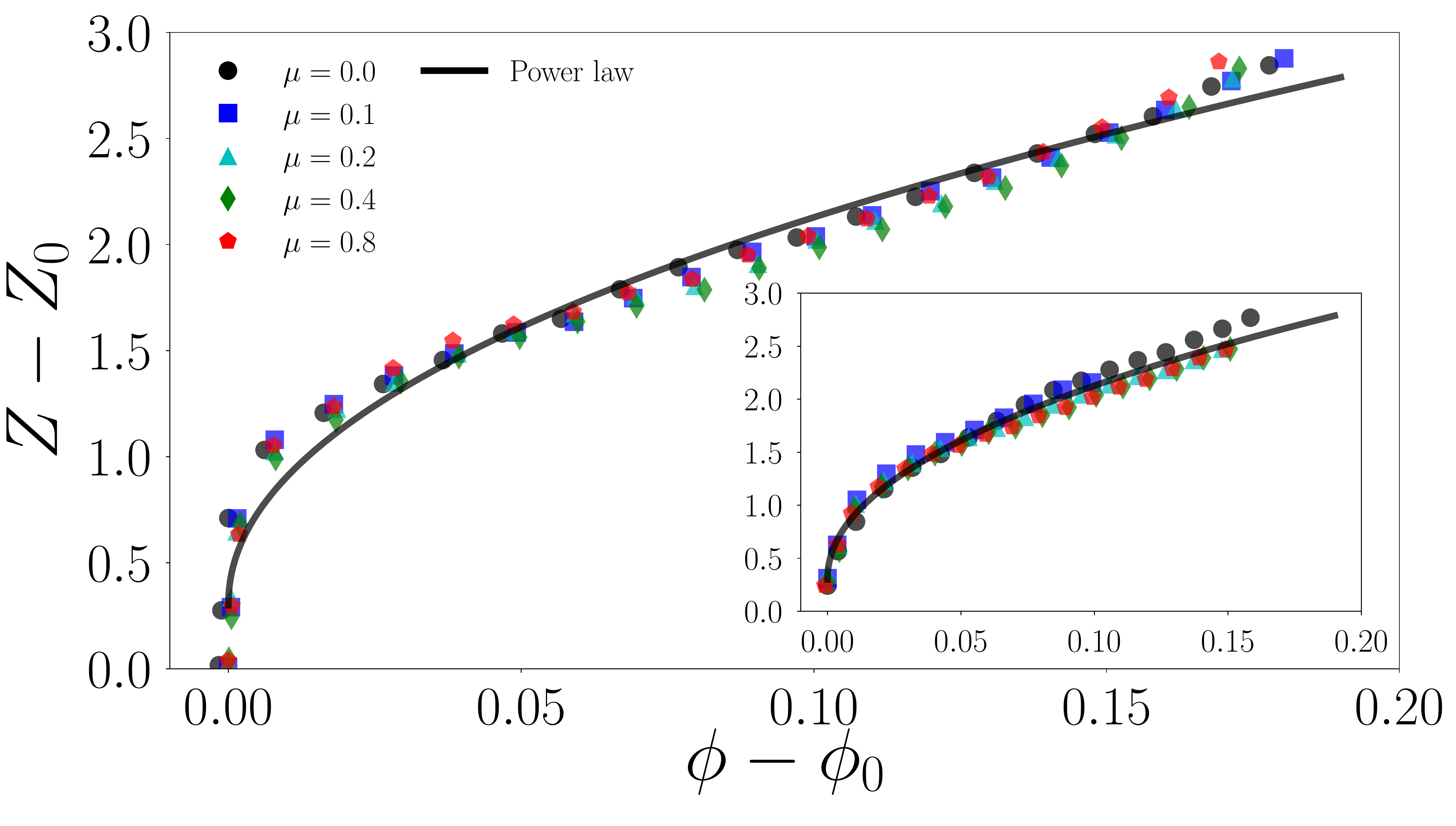}
\caption{(Color online)  Evolution of the reduced coordination number $Z-Z_0$ as a function of the reduced packing fraction $\phi - \phi_0$ for assemblies of pentagons and disks (inset), and for different values of friction coefficient.
The power-law relation  $Z-Z_0 = \xi (\phi-\phi_0)^\alpha$ with $\alpha=0.5$ and $\xi = 5.1$ is shown in a continuous line.}
\label{fig:Z_nu}
\end{figure}

Now, above the jammed state, it has been systematically reported in the literature that $Z$ continues to increase following a power-law
\begin{equation}
\label{Eq_Z_Phi}
Z-Z_0 = \xi (\phi-\phi_0)^\alpha,
\end{equation}
with $\alpha\sim0.5$, and $\xi$ a structural parameter defined as $\xi=(Z_{max}-Z_0)/(\phi_{max}-\phi_0)^\alpha$,
where $\phi_{max}$ and $Z_{max}$ are the values of $Z$ and $\phi$ when $P/E\rightarrow \infty$.
This relation has been observed in simulations and experiments for different kind of deformable systems (foams, emulsions, rubber-like particles and more recently in mixtures of rigid and deformable particles \cite{Katgert2010_Jamming,Majmudar2007_Jamming,Durian1995_Foam,Vu2019,Cardenas2020_Compaction}).
As shown in Fig. \ref{fig:Z_nu}, we found the same proportionality in our simulations, with $\xi\sim 5.1$,
independently of the shape of the particles and the friction coefficient.

\begin{figure}
\centering
\includegraphics[width=\linewidth]{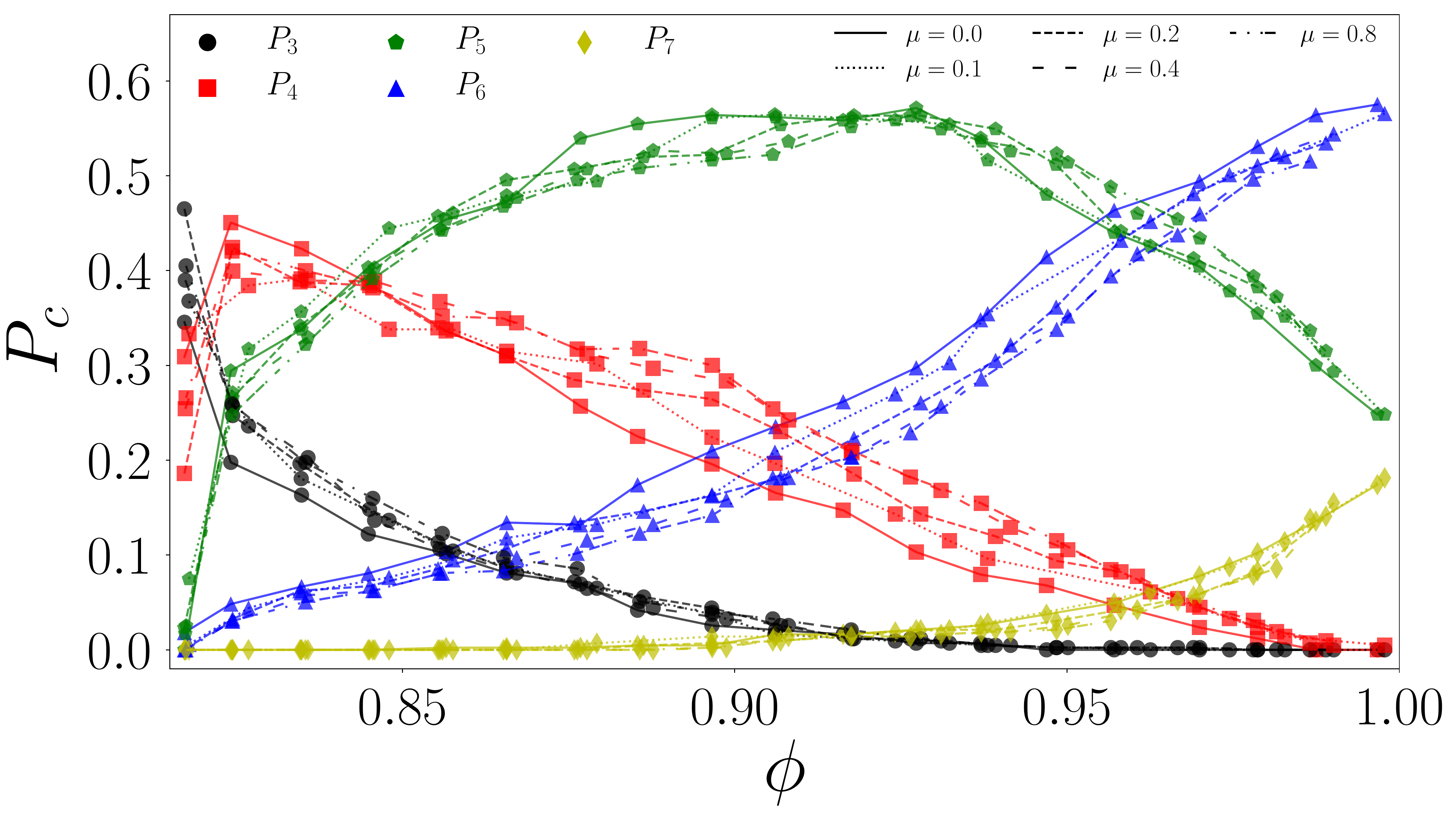}(a)
\includegraphics[width=\linewidth]{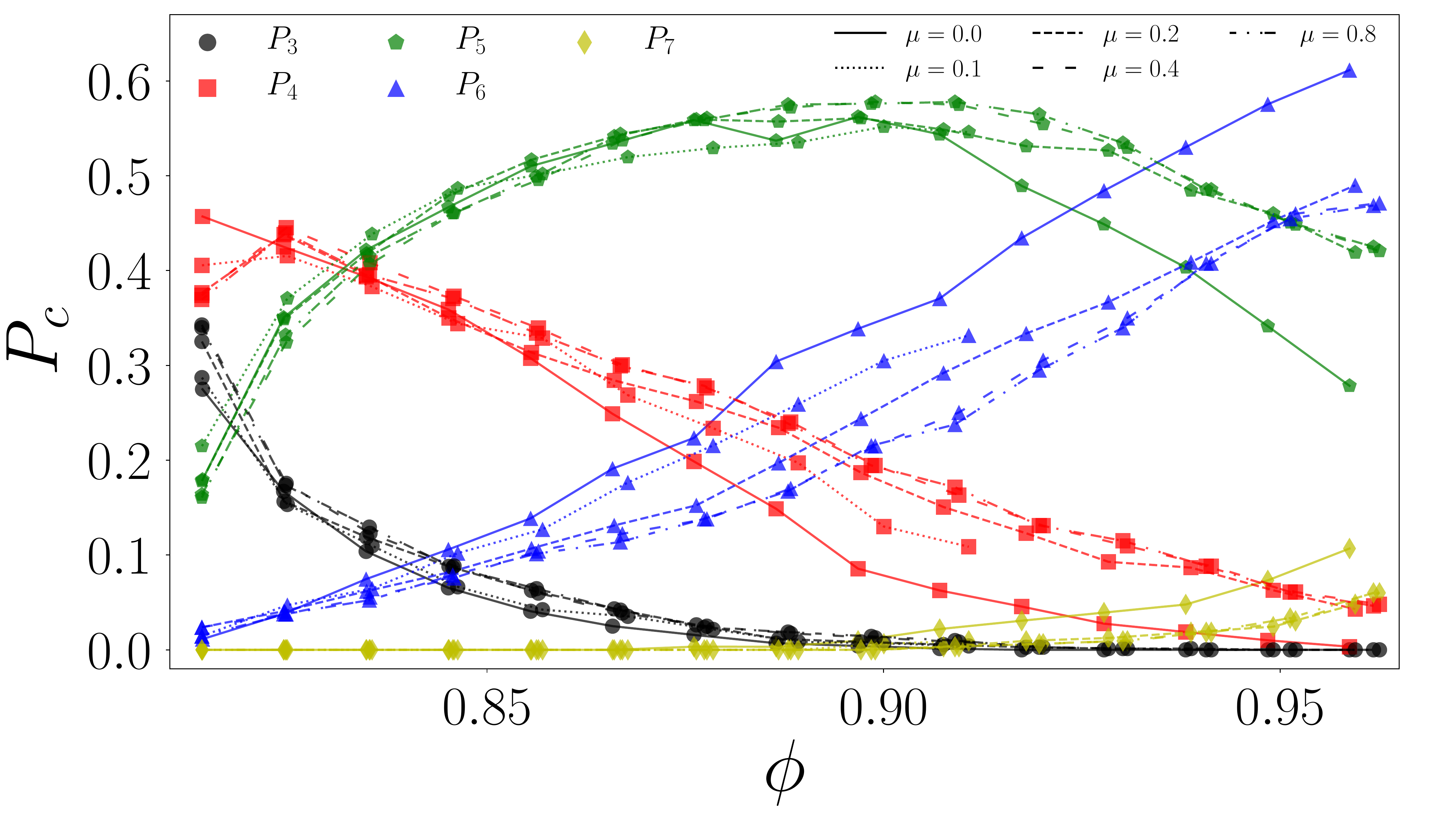}(b)
\caption{(Color online) Evolution of particle connectivity $P_c$ as a function of $\phi$:  for assemblies of (a) pentagons and (b) for discs.}
\label{fig:pc_nu}
\end{figure}

The particle connectivity can be characterized in more detail by considering $P_c$, the probability of having $c$ contacts per particle. In Fig. \ref{fig:pc_nu}, $P_c$ is plotted as a function of $\phi$ for different values of $c$.
We see that $P_c$ is nearly independent of $\mu$ for all $\phi$ values. The evolutions of $P_c$ are basically the same for both assemblies; this is,
$P_3$ and $P_4$ decreas from $\sim0.2$ and $\sim0.4$ respectively, to $0$, whereas $P_6$ increases from $0$ to $\sim 0.6$. $P_7$ increases too, but in a much slower way, from $0$ to values close to $0.1$. In contrast, $P_5$ follows a parabola with its maximum value at $\phi\sim0.92$. 

In fact, the coordination number is linked to $P_c$ by $Z = \sum_{c=1}^{\infty} c P_c$.
So, the monotonous increases of the coordination number, seen in Fig. \ref{Eq_Z_Phi}, results from
complex compensation mechanisms related to the grains' role in the contact network.
Basically, the increase in $Z$ with $\phi$ comes from the increase of $5P_5$ and $6P_6$
until $\phi\simeq0.92$ and beyond, mainly from $6P_6$ and $7P_7$.
Finally, the variations of $3P3$ and $4P4$ with $\phi$ have a minor effect on $Z$ because of the low value of $c$.

\subsection{Force and stress transmission}
\subsubsection{Force distribution}
The force chains in particle assemblies and the non-homogeneous spatial distribution of their contact forces
are topics widely studied, both numerically and experimentally. These studies are mainly performed on rigid particle assemblies of various
sizes \cite{Mueth1998_Force,Antony2000_Evolution,Nguyen2014_Effect,Oquendo2020_Densest}, shapes \cite{Donev2007_Underconstrained,Azema2009_Quasistatic,Nguyen2014_Effect} and contact interactions \cite{Valverde2006_Random,Cox2016_Self}. However, it has been seldom studied for highly deformable particle assemblies, and, in particular, in the case of non-circular deformable shapes.

\begin{figure}
\centering
\includegraphics[width=0.42\linewidth]{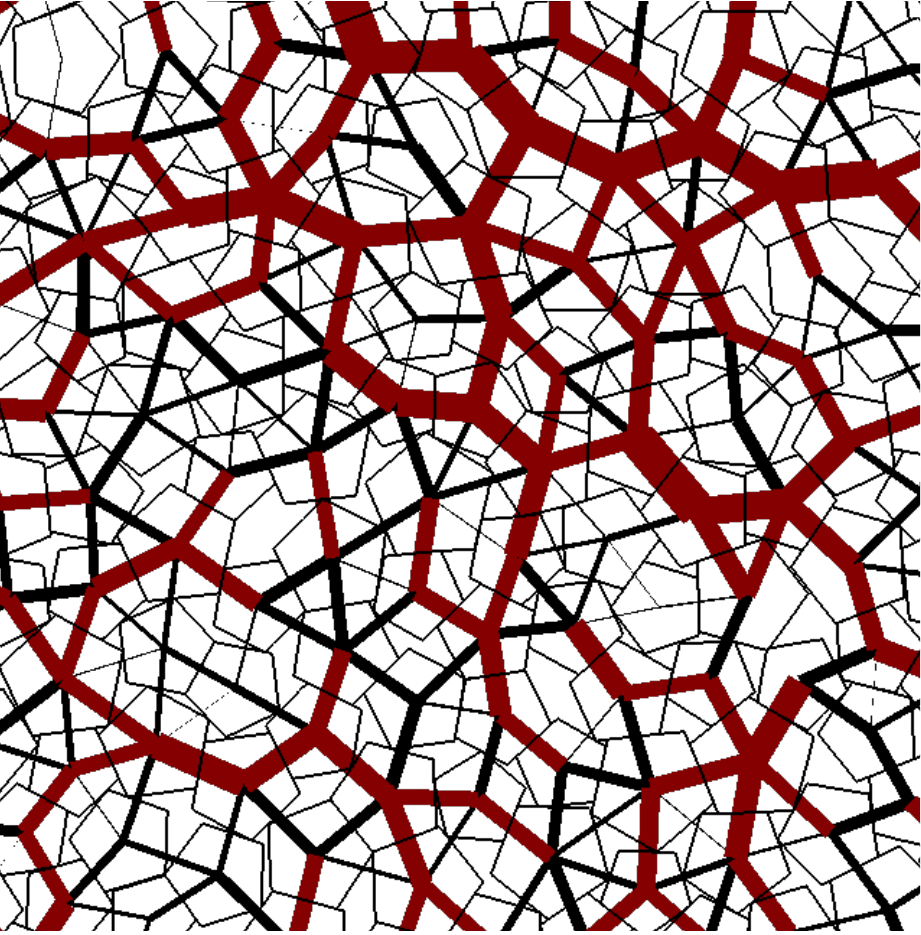}(a)
\includegraphics[width=0.42\linewidth]{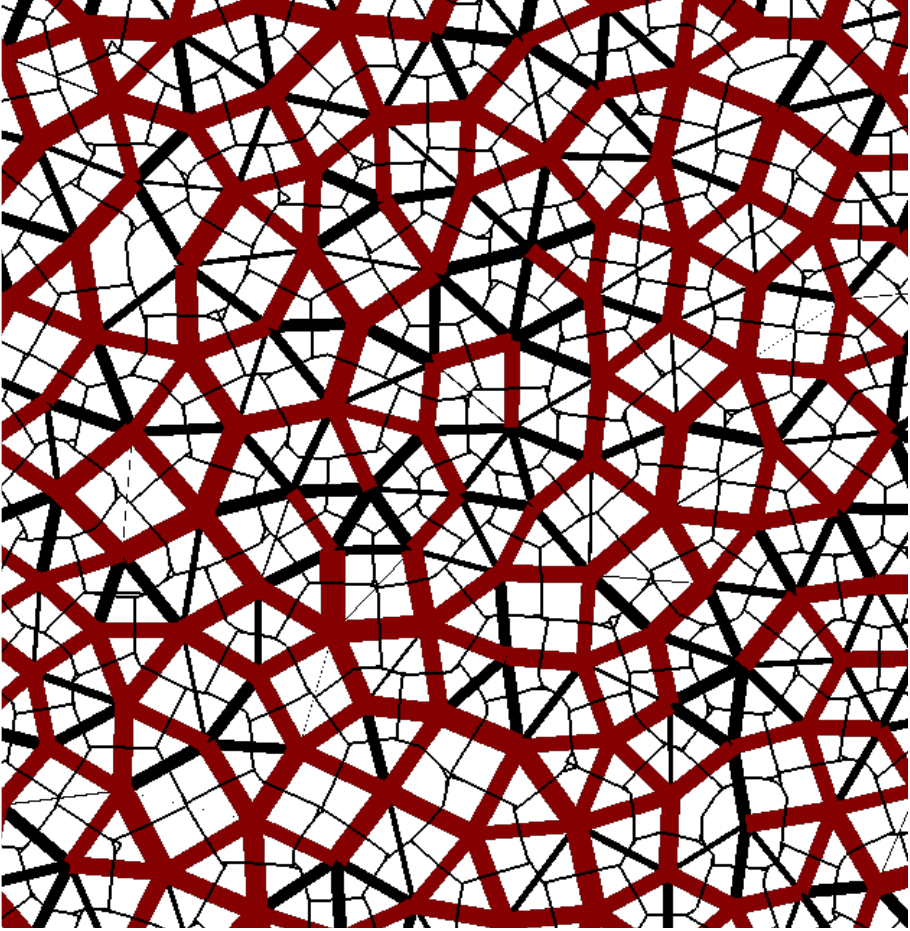}(b)
\includegraphics[width=0.42\linewidth]{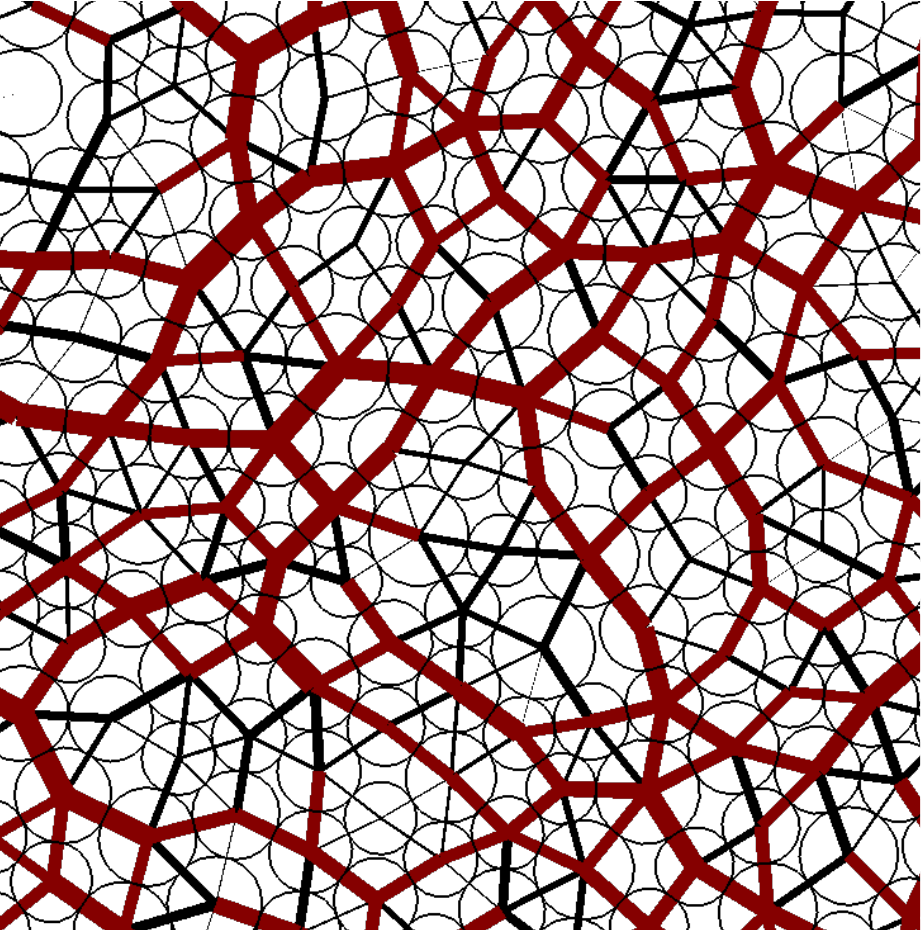}(c)
\includegraphics[width=0.42\linewidth]{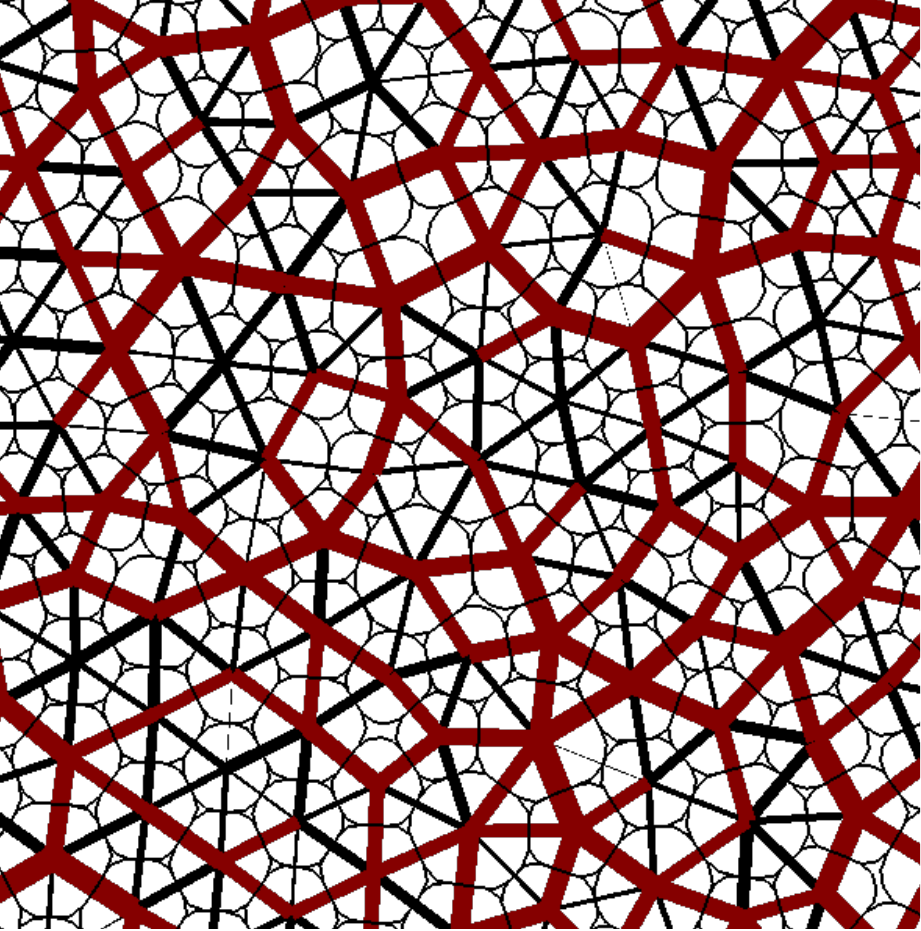}(b)
\caption{(Color online) Close-up views of the force chains in frictionless assemblies of pentagons (a,b) and  disks (c,d) at
the jammed state (a,c) and for $\phi \sim 1$ (c,d). The magnitude of each normal force is represented by the thickness of the segment joining the centers of the particles in contact. The strong forces ($f_n\geqslant\langle f_n\rangle$) and weak forces ($f_n<\langle f_n\rangle$) are plotted in red and black, respectively.}
 \label{fig:Snap-shots_forces}
 \end{figure}

Figure \ref{fig:Snap-shots_forces} shows a view of the normal forces network in assemblies of pentagons and  disks at the jammed state and for $\phi$ close to 1.
Here, the total contact force between two deformable particles is computed as the vectorial sum of the forces at the contact nodes
along the common interface (line in 2D).
Basically, the force network density (i.e., the number of force chains) increases as $\phi$ increases because, as discussed before, the mean number of contacts per particles increases. In particular, in the case of pentagons close to the jammed state, we observe stronger and more tortuous force chains, compared with disks. But, far beyond the jammed point, the force network appears to be more homogeneous in both cases.

\begin{figure}
\centering
\includegraphics[width=\linewidth]{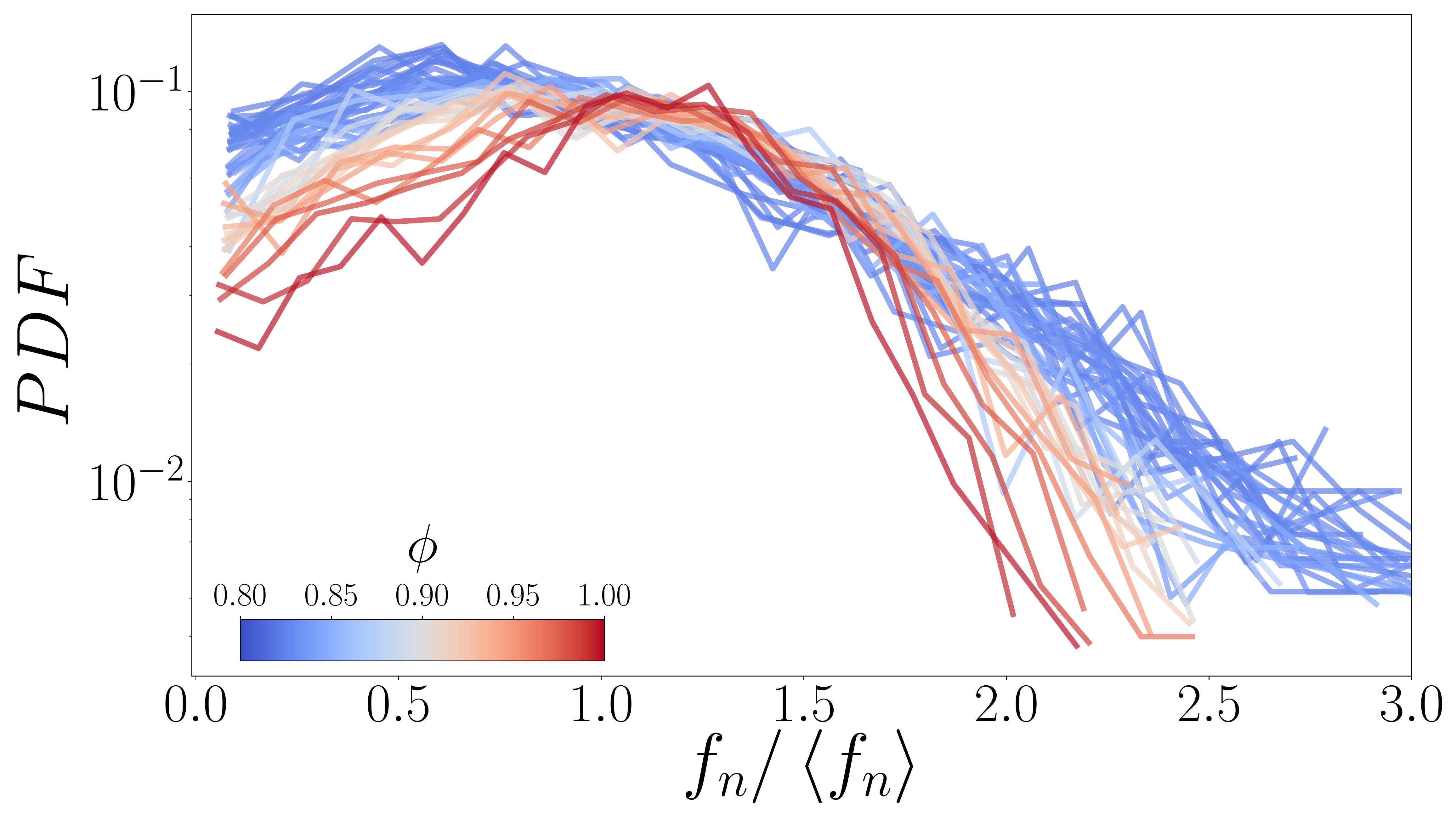}(a)
\includegraphics[width=\linewidth]{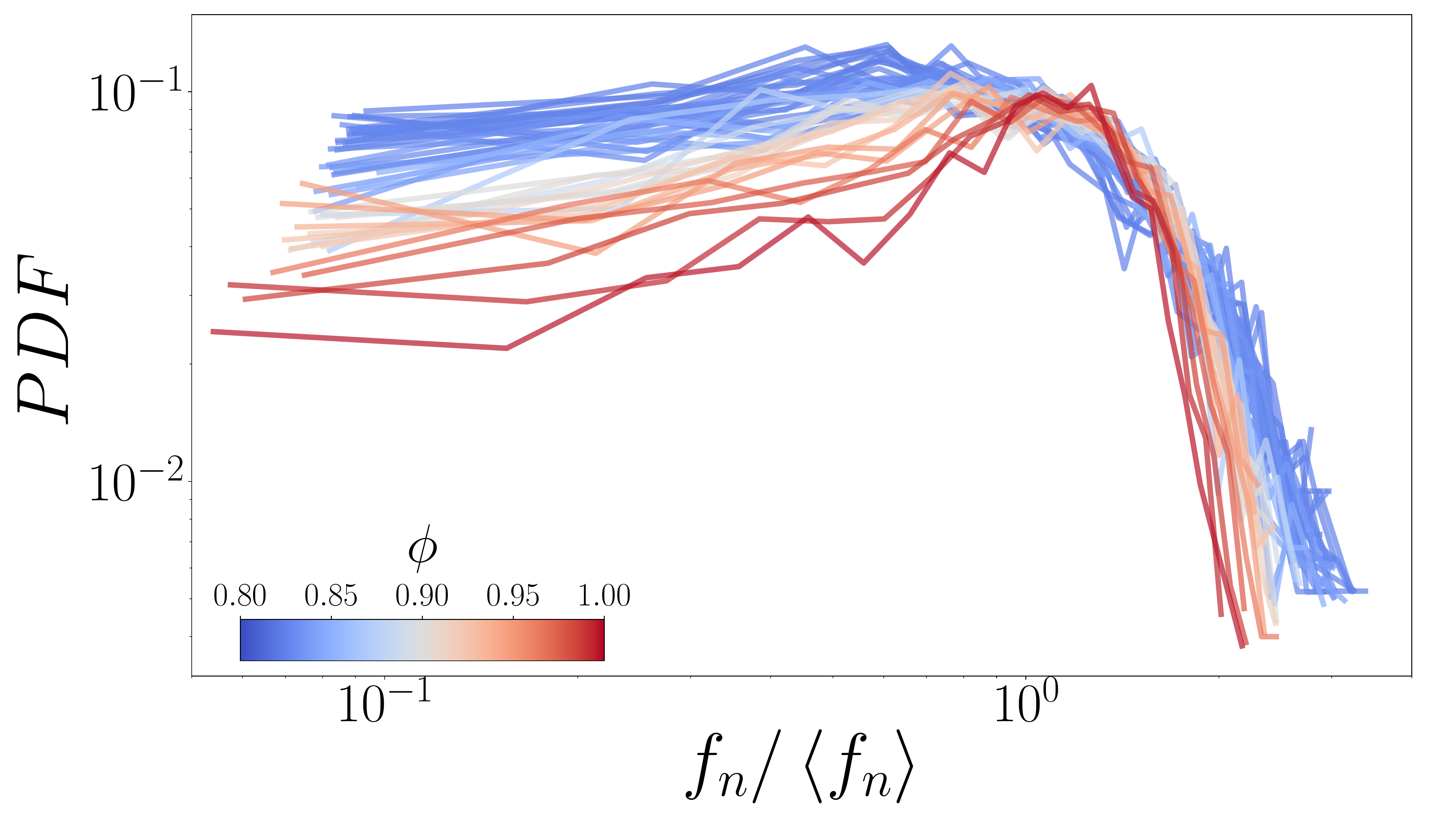}(b)
\caption{(Color online) Probability distribution function of the normal forces $f_n$ normalized by the average normal force $\langle f_n\rangle$
in (a) log-linear and (b) log-log scales for pentagons at $\mu=0$ and for different
packing fraction.}
\label{fig:pdistribution_fn}
\end{figure}

The probability density function (PDF) of the normal forces $f_n$ normalized by the mean normal force  $\langle f_n \rangle$
for frictionless assemblies of pentagons is shown in Fig. \ref{fig:pdistribution_fn}.
As is usually observed, at the jammed state the density of forces above the mean value has an exponential decay
whereas the density of forces below the mean follows a power law \cite{Radjai1998_Bimodal}.
This distribution of forces is globally maintained beyond the jammed point up to values of packing fraction close to the unity.
We also remark that the distribution becomes narrower as the packing fraction increases. Both, the maximum normal force and
the proportion of weak contacts declines as $\phi\rightarrow 1$, which implies that the force chain network becomes more homogenous.

\begin{figure}
\centering
\includegraphics[width=\linewidth]{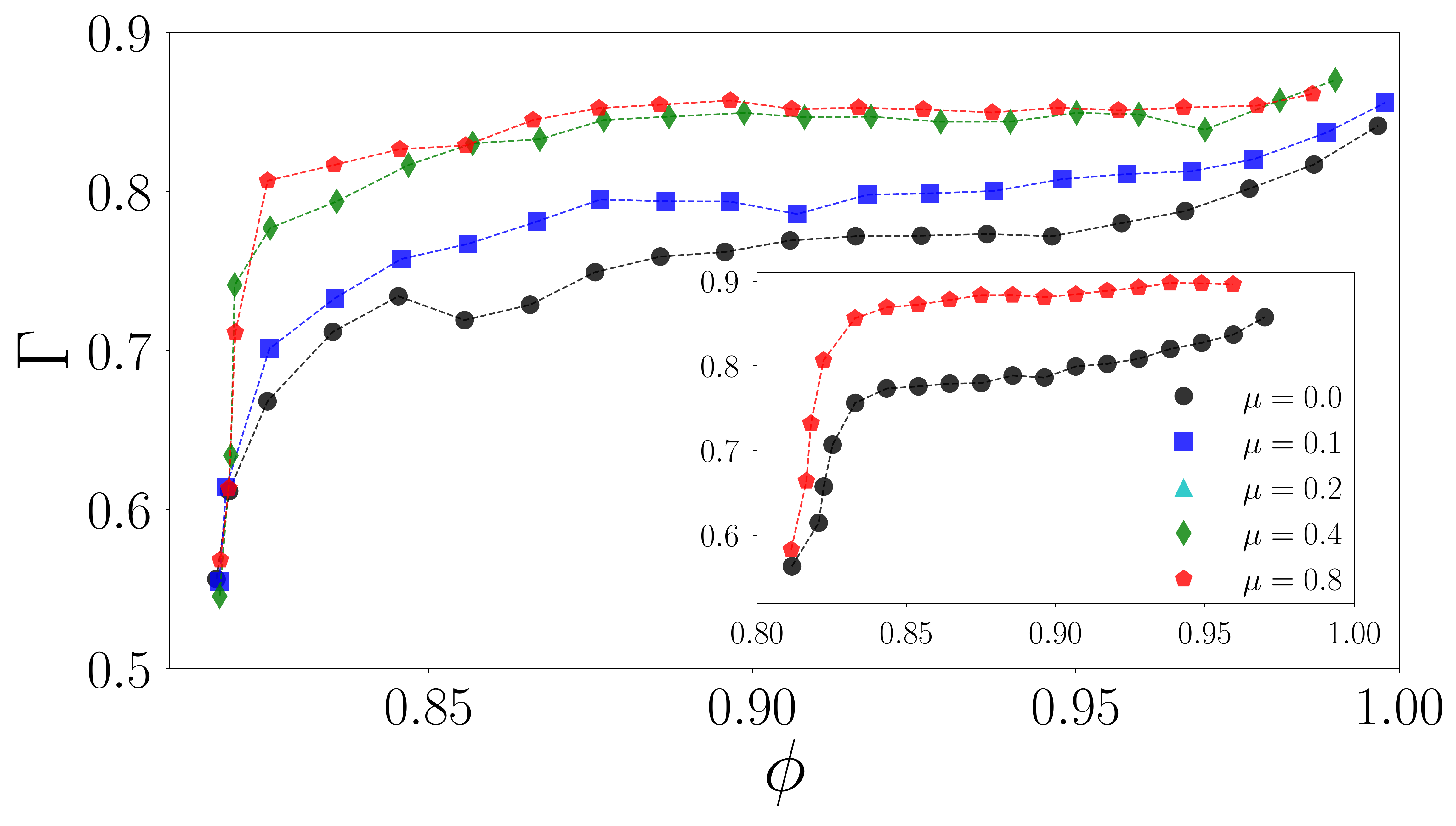}
\caption{(Color online) Participation number $\Gamma$ as a function of $\phi$ for assemblies of pentagons and disks (inset) and different friction coefficients. }
\label{fig:gamma_parameter}
\end{figure}

The degree of homogeneity of the normal force network can be quantified by the, so-called,
participation number $\Gamma$ defined as \cite{Zhang2005_Jamming}
\begin{equation}
\Gamma = \left({N_c} \sum_{i=1}^{N_c}q_i\right)^{-1},
\end{equation}
where $N_c$ is the total number of contacts in the system and $q_i = f_i/\sum_{j=1}^{N_c}f_j$, with $f_i$ the magnitude of the normal force at the contact $i$.
For a homogeneous force distribution $\Gamma$ is equal to $1$, while the limit in which the forces are completely heterogeneous corresponds to $\Gamma\simeq0$.
The evolution of $\Gamma$ as a function of $\phi$ is shown in Fig. \ref{fig:gamma_parameter}. In general, $\Gamma$ increases with $\phi$ from $\sim0.6$ at the jammed state to values close to $0.85$ at $\phi$ close to unity.
This variations of $\Gamma$ verifies that the force chain network becomes more dense and homogenous as
the packing fraction is increased far beyond the jammed point.
It is worth noting that $\Gamma$ increases with the friction, which suggests that the friction contributes to a faster homogenization of the force network.

\subsubsection{Particle stress distribution}

\begin{figure}
\centering
\includegraphics[width=\linewidth]{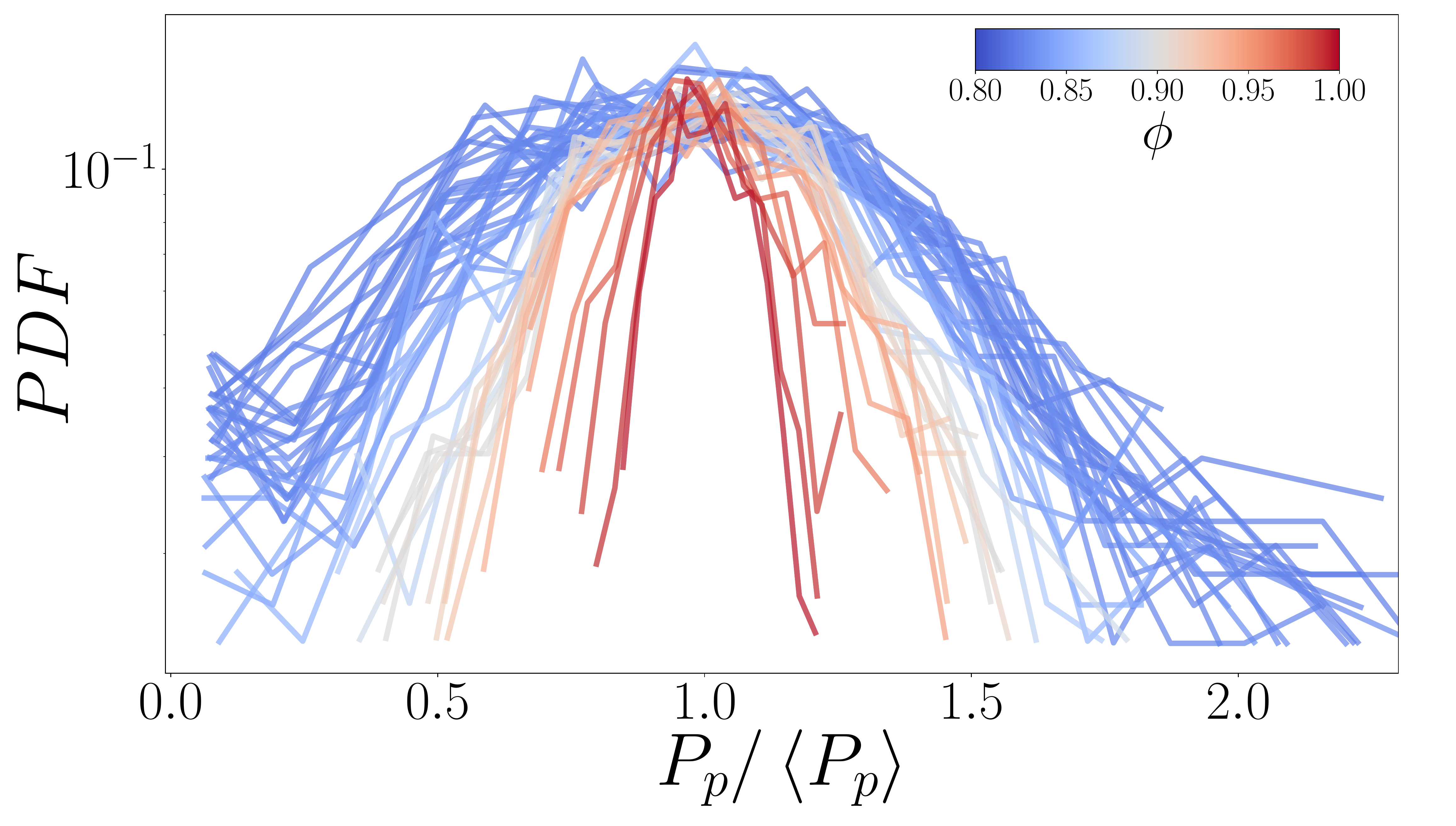}(a)
\includegraphics[width=\linewidth]{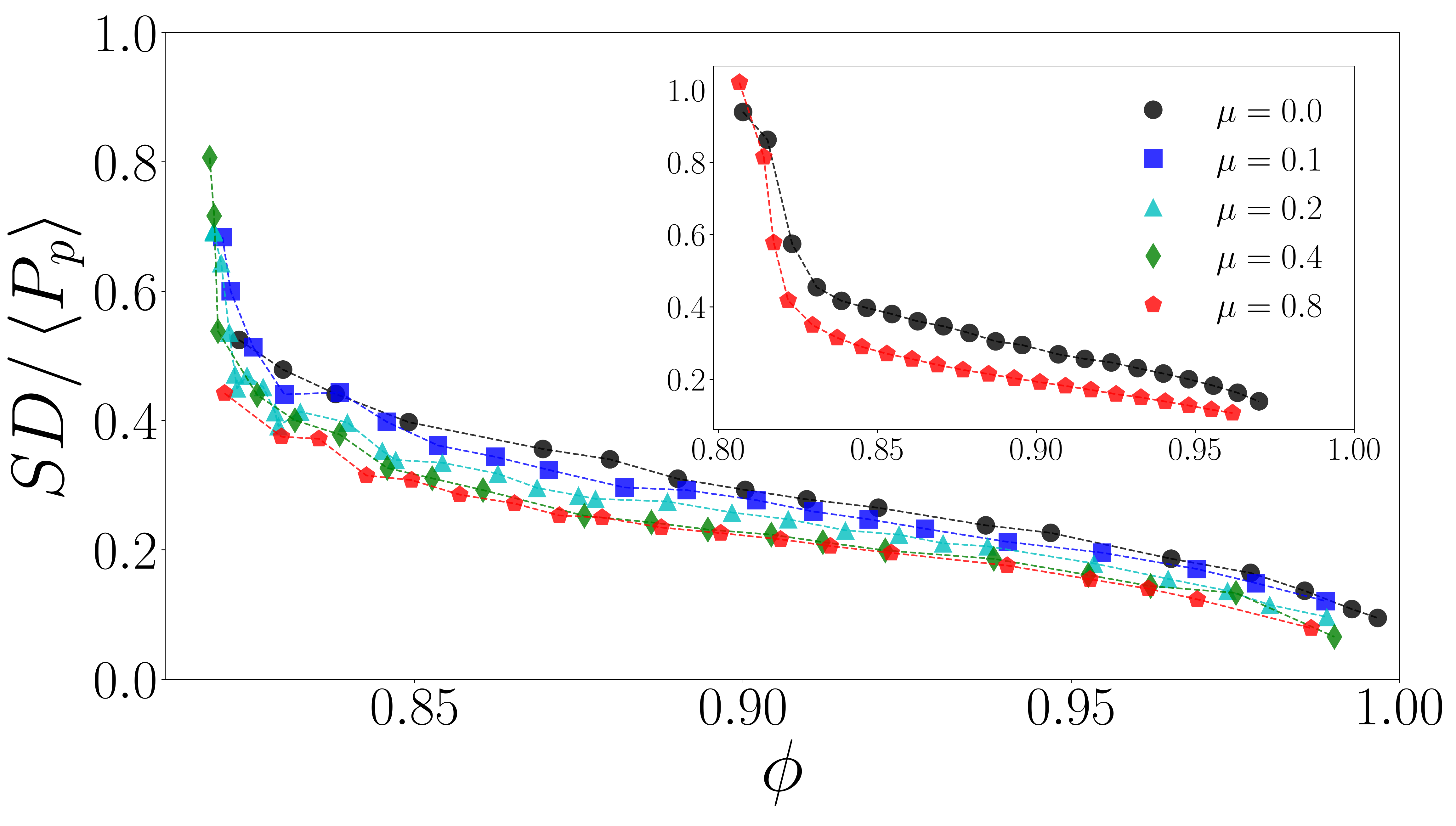}(b)
\caption{(Color online) (a) Probability distribution function of the particle stress $P_i$, normalized by the mean $\langle P_i \rangle$,
for frictionless assemblies of pentagons and increasing packing fraction $\phi$.
(b) Standard deviation of the distribution $P_p$ as a function of the packing fraction $\phi$ for
assemblies of pentagons and disks (inset) and for various values of the friction coefficient.}
\label{fig:pdf_Pp}
\end{figure}

According to the definition of the tensorial moment (Eq. (\ref{eq:M})), one can assign
to each particle $i$ a stress tensor $\bm \sigma^i = \bm M^i / V_i$.
From this particle stress tensor, we define the mean particle stress as $P_i =(\sigma^i_{1}+\sigma^i_{2})/2$, with $\sigma^i_{1}$ and
$\sigma^i_{2}$ the principal values of $\bm \sigma^i$.
The probability density function (PDF) of this particle stress, normalized by the mean $\langle P_i \rangle$, for frictionless assemblies of pentagons and for increasing packing fraction $\phi$,  is shown in Fig. \ref{fig:pdf_Pp}(a).
As a first approximation, the general shape of the distribution could be compared to a gaussian distribution around $\langle P_i \rangle$.
As the packing fraction increases the particle stress distribution narrows around the mean value,
highlighting the increasing homogenization of the stresses (in a similar way to the normal force distributions).

Figure \ref{fig:pdf_Pp}(b) shows the evolution of the relative standard deviation $SD$ of the distribution $P_p$, as a
global measure of the heterogeneities, as a function of $\phi$.
In both assemblies and for all values of inter-particles friction, $SD$ declines with
$\phi$.
Furthermore, we see that for a given value of $\phi$, $SD$ declines also with the inter-particle friction.
In other words, as observed for the distributions of forces just before (see Fig. \ref{fig:gamma_parameter}),
the particle stress also tends to be more homogeneous as the friction increases.

\section{Micro-mechanical-based constitutive equations}
\label{micro-meca-sec}

As it is mentioned above, a proper model that pictures the compaction of deformable packings should stand on its micromechanics. It means, on the physics at the scale of the particles and the contacts. In this direction, let us rewrite the granular stress tensor (Eq .\ref{eq:sigma}) from the scale of the contacts
\begin{equation}
\sigma_{\alpha \beta} =  \frac{1}{V}  \sum_{c \in V} f_{\alpha}^c \ell_{\beta}^c = n_c \langle f^c_{\alpha}\ell^c_{\beta} \rangle_c,
\label{eq:sigma_contact}
\end{equation}
where $f_{\alpha}^c$ is the ${\alpha}$-th component of the contact force acting on the contact $c$ and $\ell_{\beta}^c$ is the
${\beta}$-th component of the branch vector (the vector joining the centers of the two particles in contact).
The sum runs over all the contacts inside the volume $V$, and $\langle...\rangle_c$ is the average over all contacts.
The density of contacts $n_c$, on the right hand side of Eq. (\ref{eq:sigma_contact}), is given by $n_c=N_c/V$, with $N_c$ the total number of contacts in the volume $V$.

If we consider a small particle size distribution around the diameter $\langle d \rangle$, $\sum_{p\in V} V_p\simeq N_pV_p$,
with $V_p = (n_sd^2/8)\sin(2\pi/n_s)$, $n_s$ the number of sides of any regular polygonal particle and
the contact density can be rewritten as $n_c \simeq 4\phi Z / (n_s d^2 \sin 2\pi/n_s)$, with $Z=2N_c/N_p$, the coordination number.
From the definition of $P$ via the principal stresses of $\bm \sigma$, we get:
\begin{equation}\label{eq:Pglobal_local_contact}
P \simeq \frac{ \phi Z} {\frac{n_s}{2} \sin \frac{2\pi}{n_s}} \sigma_{\ell},
\end{equation}
with $\sigma_{\ell} = \langle f^c \cdot \ell^c \rangle_c/\langle d \rangle^2$, a measure of the mean contact stress, with $\cdot$ the scalar product.
Equation (\ref{eq:Pglobal_local_contact}) emphasizes the mutual relation between
$P$ and $\phi$ through the packing structure described by the particle connectivity $Z$ and the  contact stress ($\sigma_\ell$).

\subsection{Small deformation approach}
\begin{figure}
\centering
\includegraphics[width=\linewidth]{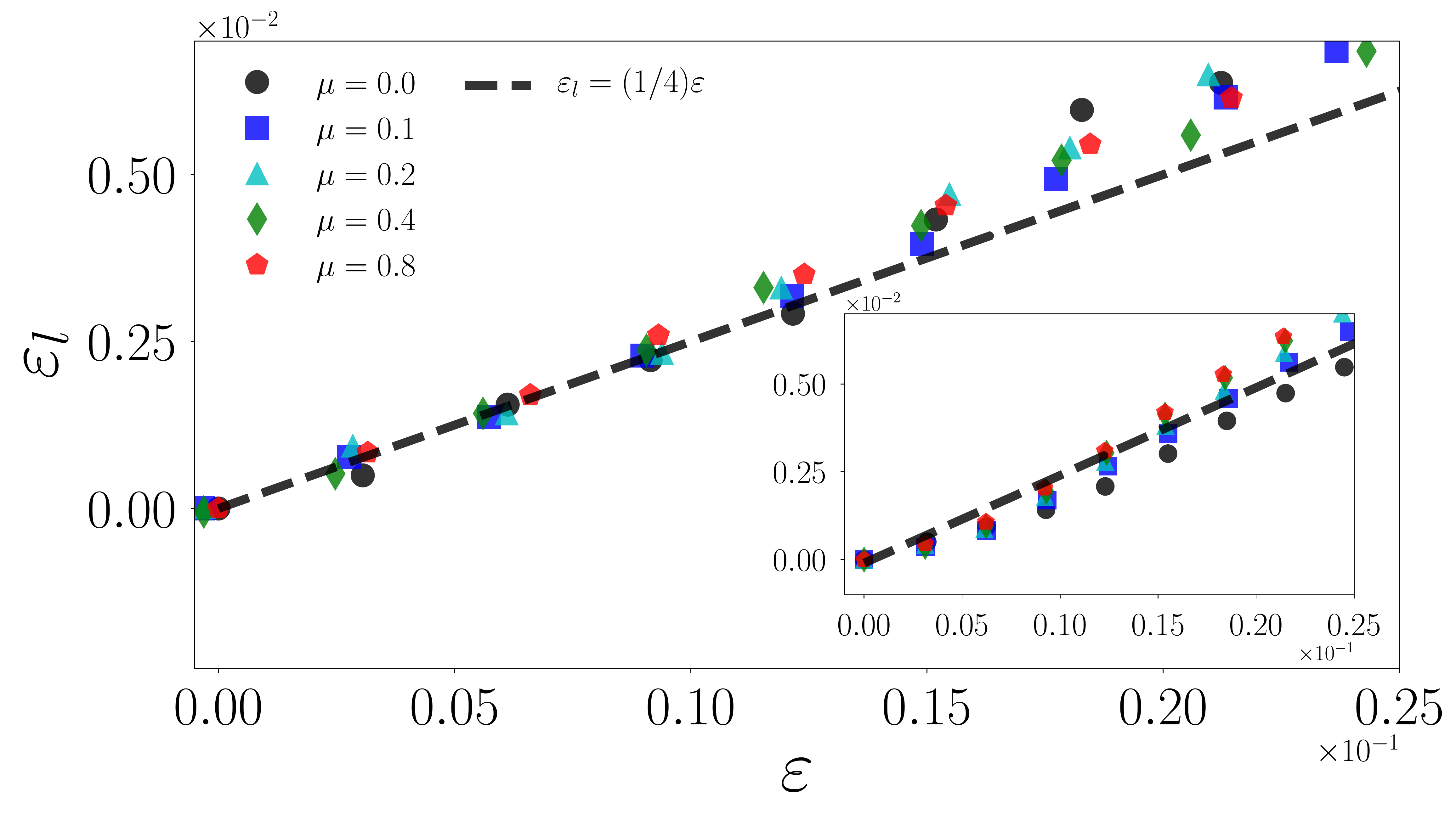}
\caption{(Color online) Macroscopic volumetric strain $\epsilon_v$ as a function of the local strain $\varepsilon_\ell$ at the
small deformation domain in assemblies of pentagons and disks (inset) for different values of friction.}
\label{fig:epsilon_epsilon}
\end{figure}

Let us consider the deformable particle assemblies as a network of bonds of length $\ell_c$,
centered on the contact point of two particles.
In the case of small and elastic deformations, we get $\sigma_\ell \sim \langle f^c\rangle_c/d = E \varepsilon_\ell$,
with $\varepsilon_\ell = \langle \ln( \ell^c /d)\rangle_c$ a local strain, defined as the mean bond strain.
Our simulations, for random packings and at small deformation, show that the local strain and the macroscopic volumetric strain are linearly dependent as $\varepsilon_\ell  \simeq (1/4) \varepsilon_v $ (Fig. \ref{fig:epsilon_epsilon}). Then, considering that $Z\rightarrow Z_0\simeq4$, Eq. (\ref{eq:Pglobal_local_contact}) leads to:
\begin{equation}
\label{eq:Pgsd}
\frac{P_{sd}}{E} = -\frac{\phi}{\frac{n_s}{2} \sin \frac{2\pi}{n_s}} \ln\left(\frac{\phi_0}{\phi}\right),
\end{equation}
the limit of $P(\phi)$ at small and elastic deformations.

The prediction given by Eq. (\ref{eq:Pgsd}) is shown in Fig. \ref{fig:pressure-phi}.
As expected, we see a fair approximation of the compaction evolution in the small-strain domain, but an increasing mismatch as the solid fraction increases.

\subsection{Large deformation approach}
In the literature, there are various theoretical approximations that relate $P$ as a function of $\phi$ for large deformations  \cite{Heckel1961,Carroll1984,Panelli2001,Secondi2002_Modelling,Zhang2014,Parilak2017,Platzer2018,Montes2018}.
All of them are based on different macroscopic assumptions that forget the micromechanical fundaments
revealed by Eq. (\ref{eq:Pglobal_local_contact}), and thus, fitting parameters are needed to adjust the proposed theoretical expressions to the data.
Nonetheless, it is relevant to note that most of them find that $P \propto \ln[(\phi_{max}-\phi)/(\phi_{max}-\phi_0) ]$.

In the micromechanical development presented here, the challenging task consists in finding a functional form for both
$\sigma_\ell(\phi)$ and $Z(\phi)$. First, for $\sigma_\ell(\phi)$ taking advantage of the proportionality of $P$ with $\phi$, together with Eq. (\ref{eq:Pglobal_local_contact}), it is easy to show that  $\sigma_\ell = -\alpha \ln[ (\phi_{max}-\phi)/(\phi_{max}-\phi_0)]$.
Where the coefficient $\alpha = (\phi_{max}-\phi_0)/(4\phi_0)$ is obtained from the limit to small deformation of Eq. (\ref{eq:Pglobal_local_contact}).
Second, we have shown that $Z(\phi)$ evolves as a power law of $\phi$ following Eq. (\ref{Eq_Z_Phi}).
Then, using the above relations, we get the final expression of $P$ as a function of $\phi$:
\begin{equation}\label{eq:Pglobal}
\frac{P}{E} = -\left(\frac{\phi_{max}-\phi_0}{4\phi_0 \frac{n_s}{2} \sin \frac{2\pi}{n_s}}\right) \{Z_0 - \xi (\phi-\phi_0)^\alpha\}
\phi \ln\left( \frac{\phi_{max}-\phi}{\phi_{max}-\phi_0} \right).
\end{equation}

Figure \ref{fig:pressure-phi} presents our numerical data for pentagons and disks (inset) assemblies together with
the compaction equation given by Eq. (\ref{eq:Pglobal}) for
$\mu_s=0$ and $\mu_s=0.8$. The predictions given by Eq. (\ref{eq:Pglobal})
are in a good agreement with our simulations, capturing the asymptotes for small and very high pressures,
the effect of the coefficient of friction and the effect of the particle shape.
In contrast to previous models,
the only unknown parameter in this new model is the maximum packing fraction $\phi_{max}$, all other constants are determined from either the initial state, the mapping between the packing fraction and the coordination number,
and the number of sides of the particles.

Going one step further, derivating Eq. (\ref{eq:Pglobal}) following Eq. (\ref{Eq_Modulus}) and neglecting small terms on $\phi \ln \phi$, we can obtain an explicit equation for the Bulk evolution
\begin{equation}\label{eq:Kglobal}
\frac{K}{E} = \left(\frac{\phi_{max}-\phi_0}{4\phi_0 \frac{n_s}{2} \sin \frac{2\pi}{n_s}}\right)  \frac{\phi^2}{(\phi_{\mathrm{max}}-\phi)}  \{Z_0 + k(\phi-\phi_0)^{\alpha}\}.
\end{equation}

Fig. \ref{fig:K-phi} shows the evolution of the above relation $K/E$ as a function of $\phi$, with a good fit for $\mu = 0.0$ and $\mu = 0.8$.

\section{Conclusions and perspectives}
\label{conclu}
In this paper, we investigate the compaction behavior of assemblies composed of soft pentagons by means of non-smooth contact dynamics simulations.
In order to see the effects of particle shape, we also simulate assemblies composed
of soft circular particles. In both cases, the deformable particles are simulated following a hyper-elastic neo-Hookean constitutive law using classical finite elements. The effect of friction was also systematically investigated by varying the coefficient of friction
from 0 to 0.8. Starting from the jammed state,
packings were isotropically compressed by applying a constant velocity on the boundaries.

As general finding, we observed that beyond the jamming state, both systems have similar behavior.
At the macroscopic scale, the packing fraction increases rapidly and tends asymptotically to a maximum value $\phi_{max}$, where the bulk modulus diverges.
At the microscopic scale, we show three important facts. First, the particle rearrangement is still important even after the jamming point. Second,
the power law relation between the coordination number and the packing fraction is still valid for assemblies of soft pentagons. And third, the contact forces and particle stress distributions become less broad as the level of compaction increases.

The main differences between the two systems come from the effect of friction.
Basically, $\phi_{max}$ declines as the friction is increased, but it decreases faster in assemblies of pentagons than in assemblies of disks.
At the micro-scale, the rearrangement of the particles is higher for soft pentagons,
although it declines as the inter-particle is increased. Interestingly, the friction between the particles also contributes to a better
homogenization of the contact force network in both systems.

Another important result is the extension of the
compaction equation previously established for soft circular particle assemblies \cite{Cantor2020_Compaction}
to soft non-circular particle assemblies.
Our model, derived from the micromechanical expression of the granular stress tensor and its limit to small and elastic deformation,
is based on the joint evolution of the particle connectivity and the contact stress.
From the expression of these well defined quantities, we establish a compaction equation and a bulk equation,
free of ad hoc parameters, perfectly fitting our numerical data. The only unknown parameter appearing in our equations is the maximum
packing fraction value.

A perspective to this work concerns the study of the effects of polydispersity of systems composed of highly deformable particles.
In this case, we expect stronger coupled effects between applied pressure, particles rearrangement, and particle size ratio
on the measured quantities. Also, an expected continuation of this work will be the analysis of the shear effects
beyond the jammed state. In particular, several questions still remain,
such as, for example, the mechanical strength and the existence of a residual state like the one observed in rigid packings.

We thank F. Dubois for the technical advice on LMGC90.
We also acknowledge the support of the High-Performance Computing Platform MESO@LR.

\bibliography{biblio}

\end{document}